\documentclass[fleqn,usenatbib]{mnras}

\usepackage{newtxtext,newtxmath}

\usepackage[T1]{fontenc}
\usepackage{ae,aecompl}

\def\be{\begin{equation}}
\def\ee{\end{equation}}
\def\CF3{{\sc cosmicflows-3}}
\usepackage{url}
\usepackage{rotating}
\usepackage{multirow}
\usepackage{verbatim}

\usepackage{graphicx}	
\usepackage{amsmath}	
\usepackage{amssymb}	

\title[2MTF and 6dFGSv power spectrum]{The Redshift-Space Momentum Power Spectrum II: measuring the growth rate from the combined 2MTF and 6dFGSv surveys}

\author[F. Qin et al.]{
Fei Qin$^{1}$\thanks{E-mail: fei.qin@research.uwa.edu.au},
Cullan Howlett$^{1}$,
Lister Staveley-Smith$^{1,2}$
\\
$^{1}$International Centre for Radio Astronomy Research (ICRAR), University of Western Australia, 35 Stirling Hwy, Crawley, WA 6009, Australia\\
$^{2}$ARC Centre of Excellence for All Sky Astrophysics in 3 Dimensions (ASTRO 3D)
}

\date{Accepted XXX. Received YYY; in original form ZZZ}

\pubyear{2015}

\begin{document}
\label{firstpage}
\pagerange{\pageref{firstpage}--\pageref{lastpage}}
\maketitle

\begin{abstract}
Measurements of the growth rate of structure, $f\sigma_8$ in the low-redshift Universe allow stringent tests of the cosmological model. In this work, we provide new constraints on $f\sigma_{8}$ at an effective redshift of $z=0.03$ using the combined density and velocity fields measured by the 2MTF and 6dFGSv surveys. We do this by applying a new estimator of the redshift-space density and momentum (density-weighted velocity) power spectra, developed in the first paper of this series, to measured redshifts and peculiar velocities from these datasets. We combine this with models of the density and momentum power spectra in the presence of complex survey geometries and with an ensemble of simulated galaxy catalogues that match the survey selection functions and galaxy bias. We use these simulations to estimate the errors on our measurements and identify possible systematics. In particular, we are able to identify and remove biases caused by the non-Gaussianity of the power spectra by applying the Box-Cox transformation to the power spectra prior to fitting. After thorough validation of our methods we recover a constraint of $f\sigma_8(z_{\mathrm{eff}}=0.03)=0.404^{+0.082}_{-0.081}$ from the combined 2MTF and 6dFGSv data. This measurement is fully consistent with the expectations of General Relativity and the $\Lambda$ Cold Dark Matter cosmological model. It is also comparable and complementary to constraints using different techniques on similar data, affirming the usefulness of our method for extracting cosmology from velocity fields.
\end{abstract} 




\section{Introduction} \label{sec:introduction}

How the large-scale structures of the Universe (the filament and void structures of the mass density field) evolve over time is one of the basic questions of cosmology, and has important implications for our understanding of gravity and the cosmological model. In particular, we seek to understand how fast this structure grows and how dark matter and galaxies are distributed in the local Universe. The first of these can be uncovered by measuring the ``growth rate of structure", defined as 
\be
f\equiv \frac{d\mathrm{ln}\,D(a)}{d\mathrm{ln}\,a}, 
\ee
where $D(a)$ is the linear growth factor and $a$ is the cosmic scale factor which is a function of time. The second of these is related to the measurements of the ``galaxy bias'' \citep{Kaiser1984,Strauss1995}, i.e. the ratio between the density of galaxies and the dark matter, defined as
\be  
b^2(\boldsymbol{k})\equiv\frac{P_{g}({\bf k})}{P_{DM}({\bf k})},
\ee
where $P_{g}$ and $P_{DM}$ are the power spectrum of the galaxies and dark matter, respectively. Usually, normalized forms of the above two cosmological parameters, written as $f\sigma_8$ and $b\sigma_8$, are considered measurable from galaxy surveys. The constant $\sigma_8$ is the root mean square (rms) of the mass fluctuation in spheres of 8 $h^{-1}$ Mpc, which is used to define the normalisation of the dark matter power spectrum.  

In past work, $f\sigma_8$ (and $b\sigma_8$) have typically been measured from the galaxy density field. For example, using the multipoles of the galaxy density power spectrum   (\citealt{feldman1994,Cole1995,Hatton1998,Cole2005,Yamamoto2006,Jennings2011,Blake2018}; and references therein), or using the two-point correlation function of the mass density field  (\citealt{Bahcall1983,Bahcall1992,Song2011,Bielby2011,Beutler2012,Howlett2015,Shi2018}; and references therein). 
Alternatively, direct measurements of the galaxy peculiar velocity field can also be used to place constraints on $f\sigma_8$. These peculiar velocities arise directly from the gravitational interactions between galaxies and the underlying density field. The use of peculiar velocity surveys to measure the growth rate and galaxy bias has a long history (see \citealt{Strauss1995} for a review) but can be broadly categorised into measurements of the velocity correlation function \citep{Gorski1989,Juszkiewicz2000,Feldman2003,Dupuy2019}, velocity power spectrum \citep{Kolatt1997,Zaroubi1997,Silberman2001,Johnson2014,Howlett2017,Huterer2017} 
or via comparison of the reconstructed/measured density and velocity fields \citep{Nusser1994,Erdogdu2006a,Lavaux2010}. 

Recent studies of the velocity fields include: \cite{Johnson2014} and \cite{Howlett2017} who used the peculiar velocity power spectrum to measure $f\sigma_8$ from the 6dFGSv \citep{Magoulas2012,Campbell2014,Springob2014} and 2MTF (\citealt{Hong2014}; Hong et. al., in prep) surveys respectively; \cite{Dupuy2019} who measured the velocity correlation function in the CosmicFlows-III dataset \citep{Tully2016}; and \cite{Carrick2015} who compared reconstruction of the 2M++ redshift compilation \citep{Lavaux2011} with peculiar velocities from SFI++ \citep{Springob2007a,Springob2007b} and the First Amendment supernova catalogue (A1, \citealt{Turnbull2012}). 

In this paper we take a different approach, using joint measurements of the galaxy density and ``momentum'' (density weighted velocity) power spectra to place new constraints on the growth rate of structure from the combined 2MTF and 6dFGSv data. This technique was first developed in \cite{Park2000} and \cite{Park2006} and is expected to contain similar information to the velocity power spectrum technique mentioned above. The comparative benefit of this technique, as shown in Howlett et. al (in preparation, hereafter referred to as Paper I), is that it can be measured and modelled in an identical manner to typical measurements of the multipoles of the galaxy density power spectrum, and is an efficient compression of the information in the velocity field compared to the techniques used in \cite{Johnson2014} and \cite{Howlett2017}. These latter methods require integration of the model power spectrum for each pair of galaxies in the data which may make them unsuitable for future, larger datasets (which is also what makes such a technique disfavoured for modern measurements of the galaxy density power spectrum). However, our method is ultimately compressing information from the full distribution of densities and velocities into a handful of measurements, and may result in a reduction in constraining power for small datasets. This is explored herein.
 
Overall, this work represents a companion paper to Paper I, and is an application of the techniques described in that work to real data. As we will demonstrate, our estimator of the momentum power spectrum allows for robust and accurate extraction of the growth rate of structure from the data even in the presence of complex survey geometry, cosmic variance and peculiar velocity errors, and non-linearities in the motions of galaxies. We include a technique to accurately accommodate the presence of non-Gaussian measurement distributions into the analysis, which is validated using mock galaxy catalogues. Comparison of our growth rate constraints to those in the literature for the same or similar datasets shows that our technique gives comparable and complementary results. 

This paper is structured as follows: in Section \ref{sec:data}, we introduce the 2MTF and 6dFGSv data and the mock catalogues we have produced to test the method and evaluate the errors on our measurements. The power spectrum estimators and models are presented in Sections \ref{PSobs} and \ref{PSmod}.
In Section \ref{pamfit}, we introduce our technique for fitting $f\sigma_8$ and $b\sigma_8$ in the presence of non-Gaussian measurement distributions which is then tested and verified in Section \ref{fittingM}. In Section \ref{fitting}, we apply this to the 2MTF and 6dFGSv surveys before comparing our growth rate and galaxy bias constraints to previous work and the predictions of General Relativity. We conclude in Section \ref{conc}. 

The fiducial cosmology adopted herein is flat $\Lambda$CDM with $\Omega_b=0.0491$, $\Omega_m=0.3121$, $\sigma_8=0.8150$ and $n_s = 0.9653$. We define the Hubble constant as $H_0=100 h$ km s$^{-1}$ Mpc$^{-1}$, with $h=0.6751$. The fiducial value of the growth rate at $z_{eff}=0.03$ is $f\sigma_8=$ 0.436.

\section{DATA AND MOCKS} \label{sec:data}

\subsection{The 2MTF and 6\lowercase{d}FGS\lowercase{v} surveys}

The combined 2MTF and 6dFGSv dataset we use in this work was introduced in detail in \cite{Qin2018a} and consists mainly of sky positions, observed redshifts and `log-distance' ratios of the redshift distances and true comoving distances of each object. `Log-distance' ratios are the preferred method for presenting peculiar velocity data as they have measurement errors which are closer to Gaussian. We use the estimator of \cite{Watkins2015} to convert these to pseudo peculiar velocity measurements which preserve Gaussianity,
as demonstrated for the 2MTF dataset by \cite{Howlett2017}.

In brief, the 2MASS Tully-Fisher survey (2MTF, \citealt{Masters2008,Hong2013,Masters2014}, Hong et al., in preparation) contains 2,062 nearby spiral galaxies in the CMB frame redshift range of $cz\in\left[600, 10000\right]$ km s$^{-1}$. The 21-cm HI rotation widths are obtained from archival data, GBT and Parkes observation and ALFALFA data \citep{Springob2005,Haynes2011,Hong2013,Masters2014}. All 2MTF photometric data and the redshift data are measured in the 2MASS Redshift Survey \citep{Huchra2012} and an apparent total K-band magnitude limit of $11.25$ mag is applied. The 2MTF data is effectively full-sky; with the only caveats being missing data within the Zone of Avoidance for Galactic latitude $|b|<5^{\circ}$, and that the projected sky density of 2MTF is greater above the declination $\delta=-40.0^{\circ}$ by a factor of 2. Both of these effects are fully accounted for in this work.

The 6dFGS peculiar-velocity survey (6dFGSv, \citealt{Campbell2014}) contains 8885 early-type galaxies with $cz < 16,200$ km~s$^{-1}$ (CMB frame). The total apparent $J$-band magnitude limit is $13.65$ mag, and velocity dispersions are cut to greater than 112 km~s$^{-1}$. Fundamental Plane fits to the data are presented in \cite{Magoulas2012}, and then converted to measurements of the log-distance ratio in \cite{Springob2014}. The sky projection of 6dFGSv is anisotropic, with galaxies limited to $\delta<0.0^{\circ}$ and Galactic latitude $|b|>10^{\circ}$. 

\begin{figure}  
\includegraphics[width=\columnwidth]{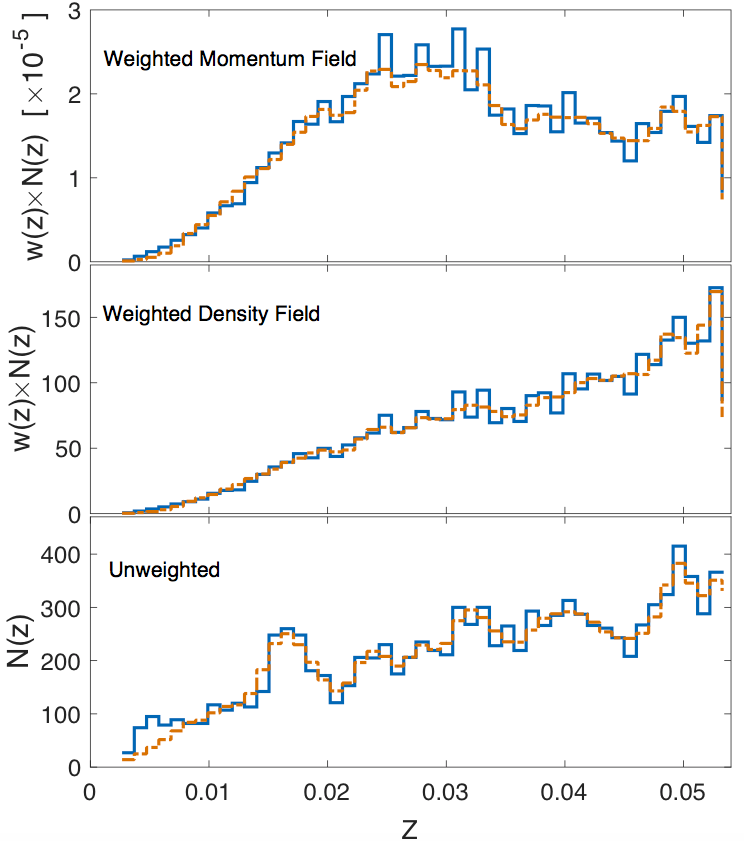}
 \caption{The redshift distribution of the combined 2MTF and 6dFGSv data and mocks (see Section~\ref{sec:mocks}). The blue solid lines indicate the distribution of the combined 2MTF and 6dFGSv data, whilst the orange dashed lines indicate the average of the combined mocks. In the top panel, the weights $w(z)$ are calculated using the first expression of Eq.\ref{wopt}, as used for measuring the momentum power spectrum. In the middle panel, $w(z)$ is calculated using the second expression of Eq.\ref{wopt}, as used for measuring the density power spectrum. The bottom panel shows the unweighted histograms of the redshifts.}
 \label{wcz}
\end{figure}

\cite{Howlett2017c} have used Fisher matrix forecasts to investigate the constraints on the growth rate, $f\sigma_8$ that can be obtained using the combined 2MTF and 6dFGSv surveys. In particular, they found that adding the 2MTF survey to the 6dFGSv survey improves the growth rate constraints by $\sim20\%$, which we will test in this work. Combining the two increases the cosmological volume compared to 2MTF alone, but adds a higher number density of objects at low redshift where the velocity errors are smaller compared to 6dFGSv. The combination gives a total of 10904 galaxies. The weighted redshift distribution of the combined data set is shown in Fig.~\ref{wcz} where the weight factors $w(z)$ used are the same as those used in calculation of the galaxy density and momentum power power spectrum (Eq.~\ref{wopt} in Section \ref{PSobs}). The weighted distributions are slightly different for the two measurements, as discussed in Section~\ref{PSobs}. The effective redshift of the combined data set is $z_{eff}\approx 0.03$.  

\subsection{The 2MTF and 6\lowercase{d}FGS\lowercase{v} mock catalogues}\label{sec:mocks}

Extracting cosmological constraints from the 2MTF and 6dFGSv data requires accurate mock galaxy catalogues. These are used to test our procedures for measuring the clustering in the data and fitting for the growth rate of structure, and to evaluate the systematic and statistical errors associated with these measurements. Following methods commonly used to measure redshift space distortions in large galaxy surveys (for example \citealt{Manera2015,Howlett2015cs} and references therein, or see \citealt{Monaco2016,Blot2019} for reviews/comparisons of different methods), we generate a large ensemble of 1000 mock surveys for each dataset, which allows us to directly estimate the covariance matrix (and cross-covariance) of the density and momentum power spectrum measurements including the effects of peculiar velocity measurement uncertainty and cosmic variance. The number of mock catalogues is chosen so that the error in our estimation of the covariance matrix remains subdominant compared to the measurement errors \citep{Percival2013}, although as we will show later we also marginalise over the uncertainty in the covariance matrix itself when fitting the data and mocks \citep{Sellentin2016}.

Producing such a large ensemble of mock surveys in a feasible time requires approximate N-Body methods and techniques to populate dark matter simulations with galaxies that match the selection functions of the data. More details about the method we use to generate samples of galaxies that mimic the properties (in particular the galaxy bias, luminosity cuts and angular/radial selection functions) of the 2MTF and 6dFGSv data are given in Appendix \ref{sec:Ap1}.

\begin{figure}  
\includegraphics[width=\columnwidth]{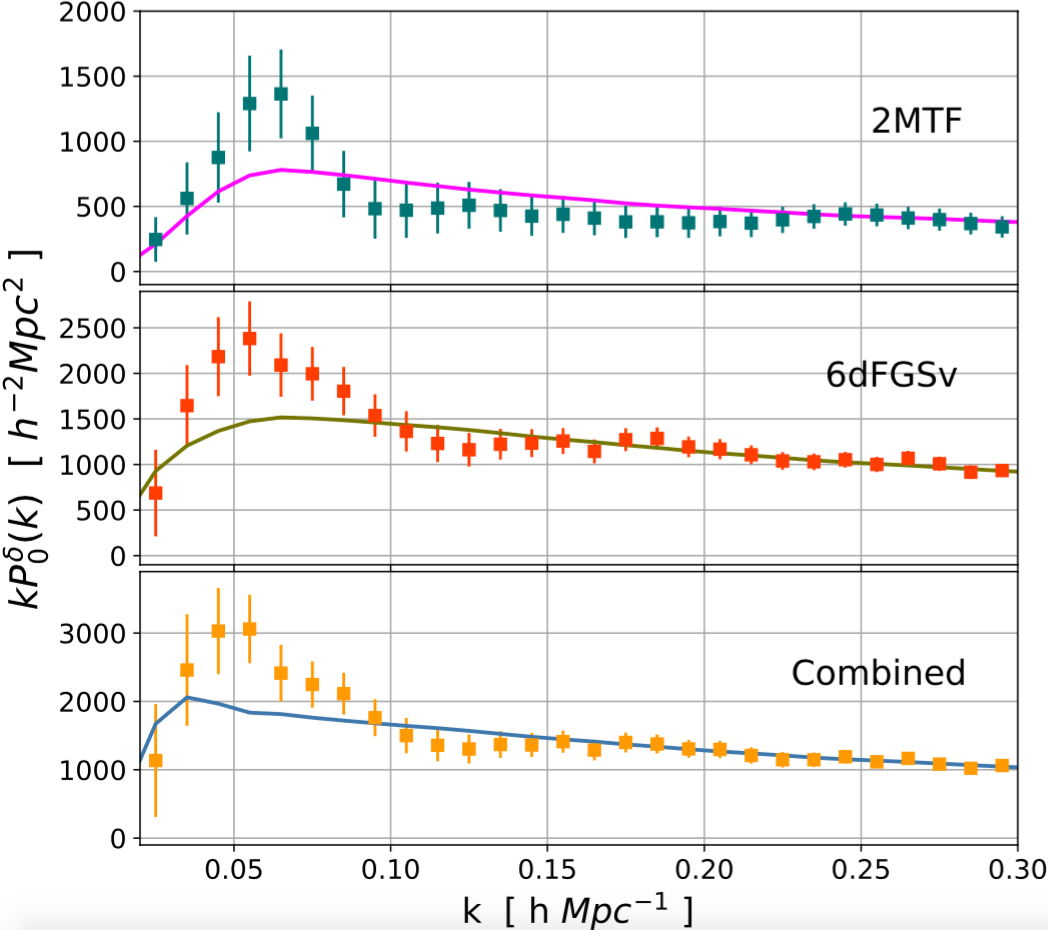}
 \caption{The density power spectra of our samples. Filled squares show the surveys. Solid lines show the mean of the mocks. The top panel is for the 2MTF, the middle panel is for the 6dFGSv, the bottom panel is for the combined sample. The $\chi^2$/d.o.f between the data and mock average on each panel are 37/25, 39/25 and 41/25, respectively.}
 \label{DAMODENs}
\end{figure}

In Fig.~\ref{wcz} we plot the weighted average number density of the combined mocks alongside that of the data. 
In Fig.~\ref{DAMODENs}, we plot the density power spectrum for the 2MTF, 6dFGSv and combined datasets alongside the average of the mocks. The density power spectrum of the individual surveys was used to tune the (three) free parameters for placing galaxies into the dark matter simulations (see Appendix~\ref{sec:Ap1}). The $\chi^2$/d.o.f between the data and mock average on each panel are 37/25, 39/25 and 41/25, respectively. In all cases we see that the mocks are in excellent agreement with the data and reproduce the selection function and clustering of the individual samples well\footnote{This is demonstrated by the fact that the values for the $\chi^{2}$ per degrees of freedom are close to one. There is significant off-diagonal covariance on large scales that makes `chi-by-eye' a poor method for estimating the goodness of the fit, and in reality the apparent poor fit at $k\approx0.05h\,\mathrm{Mpc^{-1}}$ is well within the expectations of cosmic variance.}. As such, we are confident these can be used for covariance matrix estimates of the real data.

\section{Power spectrum estimation in redshift space}\label{PSobs}

The power spectrum is the Fourier transform of the correlation function for a given field (galaxy density $\delta$, or galaxy momentum $p$). 
We start by defining the weighted field function as (\citealt{feldman1994}; Paper I)
\be\label{field}
F({\bf r})\equiv\left\{
\begin{aligned}
&\frac{w({\bf r})\left[ n({\bf r})-\alpha n_s({\bf r}) \right]}{A}~, &\delta\mathrm{-field}\\
&\frac{w({\bf r})n({\bf r})v({\bf r})}{A}~, &p\mathrm{-field}
\end{aligned}
\right.
\ee
where $w({\bf r})$, 
$n({\bf r})$ and $v({\bf r})$ are the weight, number of galaxies and line-of-sight peculiar velocity at position ${\bf r}$, respectively. Given a synthetic catalogue of random points to mimic the sky coverage and selection function of the real catalogue, $n_s({\bf r})$ is the number of the random points at position ${\bf r}$. The real catalog has a mean number density $\alpha$ times that of the synthetic catalog.
The normalization factor, $A$ is given by
\be 
A^2=\int w^2({\bf r})\bar{n}^2({\bf r}) d^3r,
\ee 
where $\bar{n}({\bf r})$ is the mean number of galaxies at ${\bf r}$. The first expression of Eq.~\ref{field} is the definition of the galaxy density field, while the second one is the definition of the galaxy momentum field as presented in Paper I. In this paper, the line-of-sight peculiar velocity $v({\bf r})$, is estimated from the measured log-distance ratios of galaxies using the estimator of \cite{Watkins2015}.

In redshift-space the measured power spectrum is anisotropic. The power spectrum $P({\bf k})$ can hence be decomposed using the Legendre polynomials \citep{Hatton1998,Yamamoto2006} and Paper I describes how the anisotropic power spectrum for the momentum field can be similarly decomposed. However, in this paper, we only measure the monopole power spectra, i.e the $l=0$ mode of the Legendre expansion as we expect higher multipoles to be completely noise dominated given the relatively small volume and number of galaxies in our sample. Following the arguments in \cite{feldman1994}, \cite{Yamamoto2006} and \cite{Bianchi2015}, the monopole power spectrum is given by
\be\label{P0k}
P_0(k)=\frac{1}{A^2}\int \frac{d\Omega_k}{4\pi}\left[ F_0({\bf k})F^*_0({\bf k}) \right]-P^{noise}_0 ,
\ee
where 
$F_0({\bf k})$ is the Fourier transform of the field function $F({\bf r})$,
\be\label{A0}
F_0({\bf k})=\int  F({\bf r}) e^{i {\bf k} \cdot {\bf r} }    d^3r,
\ee
and $F^*_0({\bf k})$ of Eq.~\ref{P0k} is the complex conjugate of $F_0({\bf k})$. The shot-noise term $P^{noise}_0$ arises from the fact that galaxies are discrete (and typically regarded as Poissonian) tracers of the density and velocity fields \citep{feldman1994,Park2000}. The expression for $P^{noise}_0$ for the density and momentum fields is slightly different and given by
\be\label{Pnoise}
P^{noise}_0=\left\{
\begin{aligned}
&(1+\alpha)\int w^2({\bf r})\bar{n}({\bf r})d^3r~, &\delta\mathrm{-field}\\
&\int w^2({\bf r})\bar{n}({\bf r})\langle v^2({\bf r}) \rangle d^3r~, &p\mathrm{-field}
\end{aligned}
\right.
\ee
where the second expression was derived in Paper I. We evaluate both of these by converting the integral to a sum over galaxies as in \cite{feldman1994}.  

In the above equations the weight factor, $w({\bf r})$ is a free parameter. 
Assuming the fluctuation of the field is Gaussian, optimal values for this weight can be found by minimizing the fractional variance of the estimated power spectrum. Performing this exercise for the monopole power spectrum gives (\citealt{feldman1994,Yamamoto2003,Yamamoto2006}; Paper I)
\be\label{wopt}
w({\bf r},k)=\left\{
\begin{aligned}
&\frac{1}{1+\bar{n}({\bf r})P^{\delta}_{0}(k)}~, &\delta\mathrm{-field}\\
&\frac{1}{\langle v^2({\bf r}) \rangle+\bar{n}({\bf r})P^{p}_{0}(k)}~, &p\mathrm{-field}
\end{aligned}
\right.
\ee
where we use the superscript $\delta$ and $p$ to distinguish between the density and momentum power spectra respectively. In practice these weights are typically implemented by choosing a constant $P^{\delta}(k)=P^{\delta}_{\mathrm{FKP}}$ that corresponds to the value of the power spectrum at the scales we wish the emphasise. We take the same approach for the momentum field. In the course of this work, we tested many different values for $P^{\delta}_{\mathrm{FKP}}$ and $P^{p}_{\mathrm{FKP}}$ when measuring the power spectra of our mock surveys and found that choosing $P^{\delta}_{\mathrm{FKP}}=1600h^{-3}\,\mathrm{Mpc^{3}} $ and $P^{p}_{\mathrm{FKP}}=5\times 10^9h^{-3}\,\mathrm{Mpc^{3}\,km^{2}\,s^{-2}}$ resulted in optimum measurements.

Finally, when weighting the momentum field, 
for measuring the momentum power spectrum 
we also require an estimate of $\langle v^2({\bf r}) \rangle$ to account for peculiar velocity measurement errors. We adopt \citep{Watkins2015,Qin2019}
\be\label{stdvs}
\langle v^2({\bf r}) \rangle=\left( \frac{\ln(10)cz}{1+z}\sigma_{\eta}(r)\right)^2+300^2\,\mathrm{km^{2}\,s^{-2}},
\ee
where $z$ is the observed redshift of the galaxies, $c$ is the speed of light, and $\sigma_{\eta}(r)$ is the error on the log-distance ratio for each galaxy. The additional term, $300^2$ km$^{2}$ s$^{-2}$ accounts for the intrinsic variance of the peculiar velocity field \citep{Jaffe1995,Sarkar2007,Feldman2010,Kashlinsky2010,Dai2011,Turnbull2012,Hong2014,Scrimgeour2016,Qin2018a}. Hence the weight for each galaxy is computed directly from its error, leading to an upweighting of galaxies with more accurate peculiar velocities. Note that the extra $300^2$ km$^{2}$ s$^{-2}$ is only used for the weighting scheme so that we do not allow very nearby galaxies to dominate the power spectrum measurements and is not included in the shot-noise calculation. The value of this intrinsic variance affects our
results only very weakly (see Appendix \ref{intri}).
 
\section{The theoretical model of the power spectrum}\label{PSmod}

To extract constraints on $f\sigma_8$ and $b\sigma_8$ we need to compare our measurements to theoretical models for the power spectra. 
The monopole power spectrum $P_0(k)$ can be given in terms of the full anisotropic model $P(k,\mu)$ by
\be\label{plkmod}
P_0(k)= \int^1_0P(k,\mu)L_0(\mu)d\mu,
\ee
where $\mu={\bf\hat{r}}\cdot {\bf\hat{k}}=\mathrm{cos}\,\theta\in[-1,1]$ is the cosine of the angle between the (line-of-sight) unit position vector ${\bf\hat{r}}$  and unit wave vector ${\bf\hat{k}}$.
 
In this paper, we use the model density and momentum power spectrum developed in Paper I. Full details of the model can be found in the appendix therein. The model density power spectrum at the effective redshift of the measurements is given by
\begin{multline}
P^{\delta}(k,\mu)=P_{00} + \mu^2(2P_{01}+P_{02}+P_{11})\\
+\mu^4(P_{03}+P_{04}+P_{12}+P_{13}+\frac{1}{4}P_{22}) ,
\end{multline}
and the model momentum power spectrum is given by
\be 
P^{p}(k,\mu)=(aH)^2k^{-2}(P_{11}+\mu^2(2P_{12}+3P_{13}+P_{22})),
\ee 
where $H$ and $a$ are the Hubble parameter and scale factor at the effective redshift. The terms $P_{mn}$ are integrals over the linear power spectrum $P_{L}$ and are fully presented in the Appendix of Paper I and in turn Appendix D of \cite{Vlah2012}. However, they are lengthy so we do not repeat them in this paper. The linear power spectrum $P_L$ is first generated at $z_{eff}=0.03$ for our fiducial cosmology using the \textsc{camb} package, then the non-linear/higher order terms are computed using this as input to the perturbation theory based model in Paper I, \cite{Vlah2012} and \cite{McDonald2009} which extends the validity of the model to smaller scales than would be possible using only the linear power spectrum. 

The $P_{mn}$ terms contain the cosmological parameters that we wish to fit. The free parameters are the growth rate $f\sigma_8$, two galaxy bias parameters $b_1\sigma_8$ and $b_2\sigma_8$ and the velocity dispersion $\sigma_v$. $b_1$ and $b_2$ are the linear and non-linear galaxy bias under the Eulerian bias expansion of \cite{McDonald2009}, respectively.\footnote{In the full expansion there are actually four distinct bias terms, however the other two terms in addition to $b_{1}$ and $b_{2}$ can be written as a function of $b_{1}$ \citep{Saito2014}}. The damping parameter $\sigma_v$ accounts for non-linear galaxy motions. The model in Paper I advocates for two separate velocity dispersion parameters that enter in different $P_{mn}$ terms. This is necessary for the simulations in their work, but the smaller number objects and larger peculiar velocity errors in the 2MTF and 6dFGSv data means we only require a single parameter.   
In this paper, we focus mainly on $f\sigma_8$ and $b_1\sigma_8$, as the other parameters remain effectively unconstrained \footnote{$b_2$ only enters on non-linear, noise-dominated scales for both the density and momentum power spectrum, hence it is difficult to obtain a good constraint on this parameter. As such, we treat this purely as a nuisance parameter and still allow it to vary, but only treat the linear term $b_1$ as a parameter of interest.} (but must be left free to ensure the parameters of interest are unbiased). 

To compare the model power spectrum to the measured power spectrum, we also need to account for the survey geometry by convolving the model power spectrum with the Fourier transform of the survey window function. In this work, we follow the procedure presented in \cite{Ross2013} (Section 3.3 and Appendix A therein) to calculate the convolution. The convolved model power spectrum in the $i$-th $k$-bin is given by
\be  
P^c_0(k_i)=\sum_jW[k_i][k_j]P_0(k_j)-P_{IC}P_r(k_i),
\ee  
where $P_{0}(k_j)$ is the model power spectrum in the $j$-th $k$-bin calculated from Eq.~\ref{plkmod}. $P_r(k_i)$ is the power spectrum of the window function itself, computed from the synthetic random catalogue which matches the survey geometry. $P_{IC}$ is given by
\be 
P_{IC}=\sum_j W[0][k_j]P_0(k_j)/P_r(0).
\ee 
The window function matrix $W[k_i][k_j]$ is given by
\be 
W[k_i][k_j]=\int\int P_r(x)x^2\Theta(r_x,k_j)dx d\mu
\ee  
where $r_x=\sqrt{k^2_i+x^2-2k_ix\mu}$ and $\Theta(r_x,k_j)$ is one if $r_x$ lies in the $j$-th $k$-bin and zero elsewhere. 

Although we do not need a random catalogue for computing the momentum power spectrum, we do need one for estimating the window function and we must include weights for the random points that match the weights assigned to the data. However, as described in Section~\ref{PSobs}, we allow these weights to depend on the error on each galaxy. For the optimal weights of the random catalogues used in the calculation of $P_r$ we instead adopt $\langle v^2({\bf r}) \rangle=(0.177H_0d_z)^2+300^2$ km$^2$ s$^{-2}$ for 2MTF randoms \citep{Qin2018a} and $\langle v^2({\bf r}) \rangle=(0.324H_0d_z)^2+300^2$ km$^2$ s$^{-2}$ for 6dFGSv randoms \citep{Scrimgeour2016}. This provides a good fit to average variance of the measurements in these two surveys and allows us to accurately compute the weighted survey geometry.

\section{Fitting method}\label{pamfit}

\begin{figure*} 
\includegraphics[width=177mm]{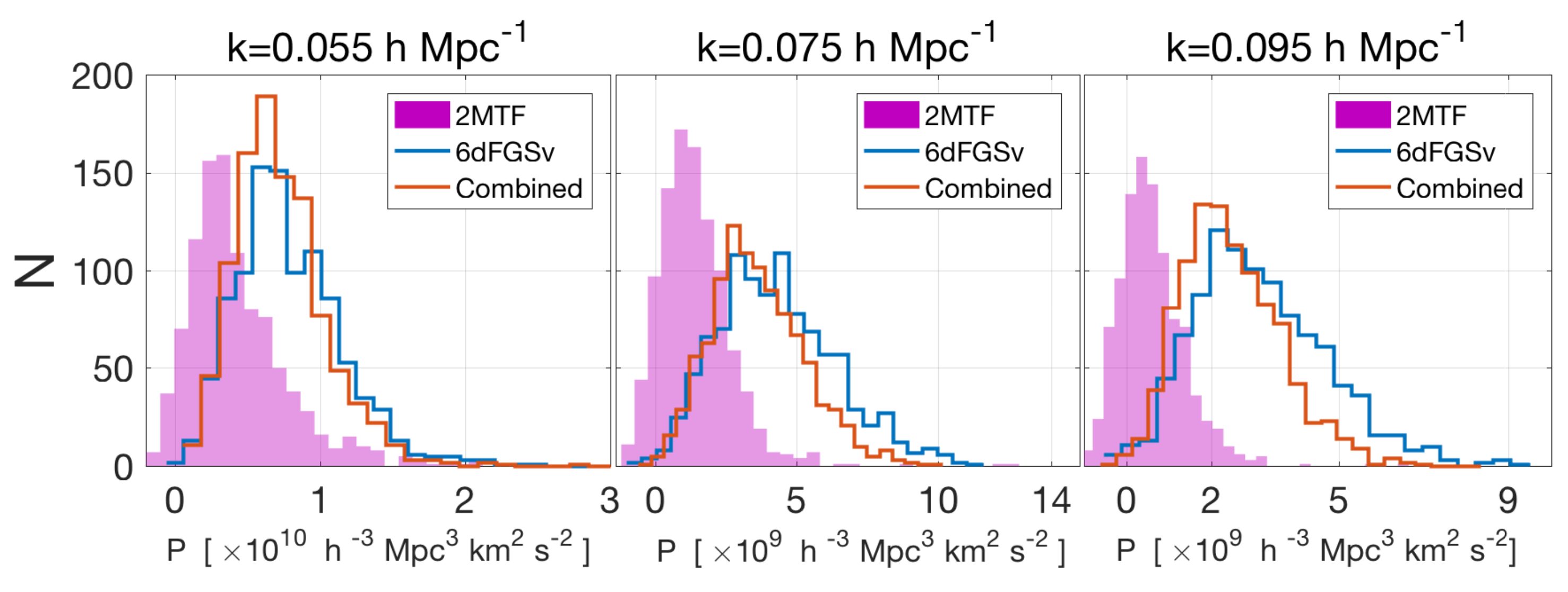}
\caption{The distribution of the momentum power spectrum measured at $k$= 0.055, 0.075, 0.095 h Mpc$^{-1}$ using mocks. The pink bars are for the 2MTF mocks, the blue lines are for the 6dFGSv mocks, the red lines are for the combined mocks.}
\label{phist}
\end{figure*}

$\chi^2$ minimization is a well-known method for parameter estimation. However, it only returns an unbiased estimation of parameters if the sample has Gaussian errors. In analysis of galaxy surveys, it is usually assumed (and motivated by the central-limit theorem) that the measured power spectrum is Gaussian distributed. Recent work \citep{Kalus2016,wang2018} has looked at this assumption in more detail and found that it breaks down on scales approaching the size of the survey volume as the number of modes available becomes small and the central limit theorem no longer holds. In addition, Paper I showed that the fact the momentum power spectrum contains a convolution between the density and velocity fields leads to intrinsic non-Gaussianity in the measurement. The fact that peculiar velocity measurement errors (using the \citealt{Watkins2015} estimator) are Gaussian and dominate on small scales alleviates this problem substantially. However we found that significant non-Gaussianity remains in the measured momentum power spectrum of the 2MTF mocks and data. This is shown in Fig.~\ref{phist}. As presented in \cite{Magoulas2012}, \cite{Scrimgeour2016}, \cite{Howlett2017} and \cite{Qin2018a,Qin2019}, the mock sampling algorithms for 2MTF and 6dFGSv are well developed, and accurately reproduce the selection functions, survey geometry and clustering of the real data. Therefore, we expect the 2MTF and 6dFGSv mocks to accurately represent the underlying distribution from  which the data is drawn. If non-Gaussianity exists in the distribution of the mocks, we believe real data should also share the same non-Gaussianity. We found this issue was much less prevalent in the 6dFGSv data/mocks and the mocks in Paper I, likely due to the larger redshift range (leading to more measurable modes) in these data combined with the relatively small errors of 2MTF.

Following \cite{wang2018} we deal with this non-Gaussianity by applying the Box-Cox transformation \citep{Box1964,Sakia1992} to the power spectrum, denoted here as $P$, to obtain the Gaussianized power spectrum, $Z$. The transformation is defined as \citep{Box1964}
\be\label{boxc1}
Z\equiv \frac{P^{\nu}-1}{\nu}~,~~\nu \in (0,+\infty).
\ee
Mathematically, the range of $\nu$ can be $(-\infty,~+\infty)$ (at $\nu=0$, $Z\equiv \mathrm{log}_{10}P$). However, following the argument in \cite{wang2018}, the range for $\nu$ should be chosen as $(0,+\infty)$ in order to ensure regularity \footnote{Infact, for our three datasets, the best estimated values of $\nu$ are all in the interval of $\nu\in(0,1)$.}.

Many methods have been developed to estimate the transformation parameter $\nu$ \citep{Box1964,CARROL1982a,SWEETING1984,Taylor1985}. For simplicity, in this paper, we follow the steps presented in Section 3 of \cite{Box1964}, using maximum likelihood estimation to estimate $\nu$. The logarithmic likelihood function of $\nu$ is given by 
\be\label{boxclog}
L(\nu) \sim (\nu-1)\sum_i^N\mathrm{log}_{10}(P_i)-\frac{N}{2}\mathrm{log}_{10}\left( \frac{\sum_i(Z_i(\nu)-\bar{Z}(\nu))^2}{N}   \right),
\ee
where $N$ is the number of independent mocks used in the power spectrum measurements. To reiterate the estimation procedure of $\nu$ presented in \cite{Box1964}, the power spectrum is measured in $N_k$ $k$-bins and in each $k$-bin, we perform the following steps:
\begin{enumerate}
\item{For a given $\nu$ in $(0,+\infty)$, use Eq. \ref{boxc1} to convert $P_i$ to $Z_i(\nu)$  ($i=1,2,...,N$).}
\item{Substitute $P_i$, $Z_i(\nu)$ and $\nu$ into Eq.~\ref{boxclog} to calculate $L(\nu)$.}
\item{Repeat the above two steps to find the value of $\nu$ in $(0,+\infty)$ that maximizes $L(\nu)$.}
\end{enumerate}
Following this, we obtain $N_k$ estimated $\nu$ values to transform the non-Gaussian power spectrum measured in $N_k$ $k$-bins. 

\begin{figure*} 
\includegraphics[height=100mm,width=177mm]{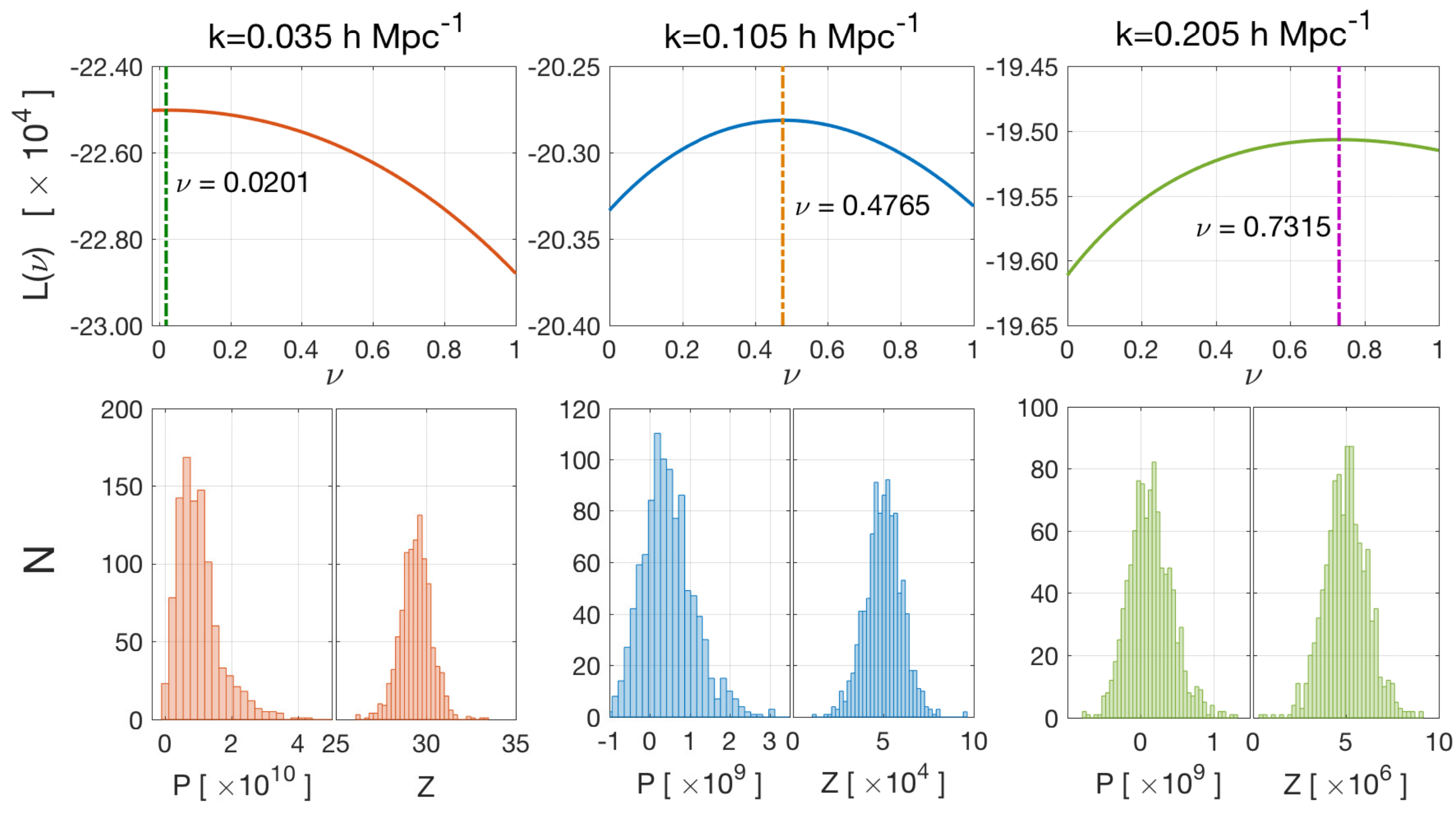}
\caption{The Box-Cox transformation of the momentum power spectrum measured from 1000 2MTF mocks at $k$=0.035, 0.105 and 0.205 h Mpc$^{-1}$. The red, blue and green curves in the three top panels show the logarithmic likelihoods $L(\nu)$ for each $k$-bin with the maximum likelihood highlighted by the vertical line. In the bottom panels, the histograms show the distributions of the power spectrum before and after the Box-Cox transformation is applied; $Z=(P^{\nu}-1)/\nu$, where $\nu=$0.0201, 0.4756 and 0.7315 for $k$=0.035, 0.105 and 0.205 h Mpc$^{-1}$, respectively. }
\label{BCnu}
\end{figure*}

As an example, Fig.~\ref{BCnu} shows the result of applying the Box-Cox transformation to the momentum power spectrum measured from $N$=1000 2MTF mocks at $k$=0.035, 0.105 and 0.205$h\,\mathrm{Mpc^{-1}}$. We show the histograms of the measured and Gaussianized power spectra as well as the likelihood $L(\nu)$, highlighting the maximum value. Applying the Box-Cox transformation with the best-fit value of $\nu$ significantly reduces the non-Gaussianity in the distribution of the measurements. As the value of $k$ increases, the distribution of $P$ tends to be more Gaussian and therefore the best-fit transformation parameter $\nu$ moves closer to 1.          

We applied the Box-Cox transformation to all three data sets, although compared to 2MTF the intrinsic distributions of the power spectra of the 6dFGSv mocks and the combined mocks are much more Gaussian\footnote{We found that for the 6dFGSv mocks and the combined mocks, the $\chi^2$ minimization returns unbiased fits for $f\sigma_8$ even without the Box-Cox transformation.}. We thus standardize the fitting method for all three data sets and better account for non-Gaussianity in all of them. 

Given a method for Gaussianizing the power spectra (Eq.~\ref{boxc1}), we can then fit the data with our model power spectrum and extract the cosmological parameters.
Denoting the Gaussianized measured and convolved model power spectra as ${\bf Z}_{o}$ and ${\bf Z}^{c}_{m}$ respectively we can use the $\chi^2$ to evaluate the likelihood of the data given the cosmological parameters;
\begin{multline}\label{chi2}
\chi^2(\boldsymbol{Z}_{o}|f\sigma_{8},b_{1}\sigma_{8},\boldsymbol{\theta})= \\
({\bf Z}_{o}-{\bf Z}^{c}_{m}(f\sigma_{8},b_{1}\sigma_{8},\boldsymbol{\theta}))\boldsymbol{\mathsf{C}}^{-1}({\bf Z}_{o}-{\bf Z}^{c}_{m}(f\sigma_{8},b_{1}\sigma_{8},\boldsymbol{\theta}))^T,
\end{multline}
where $\boldsymbol{\theta}$ denotes the parameters we consider as nuisance parameters ($b_{2}\sigma_{8}$,$\sigma_{v}$) and $\boldsymbol{\mathsf{C}}^{-1}$ is the covariance matrix of the Gaussianized power spectra. As we are fitting both the density and momentum power spectra simultaneously, both with $N_{fit}$ measurement bins (the value of $N_{fit}$ is given in Section \ref{fittingM}), the covariance matrix is a $(2N_{fit})\times (2N_{fit})$ matrix.

Finally, the true covariance matrix of the Universe is unknown; $\mathsf{C}^{-1}$ of Eq.~\ref{chi2} is just an estimated value using mocks. A fully Bayesian treatment should therefore take into account the uncertainty in the estimation of $\mathsf{C}^{-1}$ itself. \cite{Sellentin2016} show that the correct method to account for this requires using a revised $t$-distribution to calculate the likelihood of the data given the model,        
\be\label{tdis}
\mathcal{L}(Z|f\sigma_{8},b_{1}\sigma_{8},\boldsymbol{\theta})=c_p|\mathsf{C}|^{-\frac{1}{2}} \left[  1+\frac{ \chi^2(Z|f\sigma_{8},b_{1}\sigma_{8},\boldsymbol{\theta})}{N-1}  \right]^{-N/2},
\ee
where the normalization constant, $c_p$ is given by
\be\label{tdisnor}
c_p=[\pi(N-1)]^{-N_k/2} \frac{\Gamma(N/2)}{\Gamma(N/2-N_k/2)}  
\ee
and $\Gamma$ is the Gamma function.

Given the above likelihood, we use the Metropolis-Hastings Markov-chain Monte Carlo (MCMC) algorithm to recover the posterior of the cosmological parameters given the observed density and momentum power spectra. In computing the posterior we use flat priors in the interval $f\sigma_8\in(0,2]$, $b_1\sigma_8\in[0.2,5.0]$, $b_{2}\sigma_{8}\in[-2,10]$ and $\sigma_{v}\in(0,100]$ Mpc h$^{-1}$. In our theoretical model of the power spectrum, $\sigma_v$ is scaled by the inverse of the Hubble parameter and so whilst strictly a velocity dispersion, we treat it as though it has units of Mpc h$^{-1}$ (see Paper I).

\section{Tests on simulations}\label{fittingM}

To see how well the power spectrum estimation method (presented in Section \ref{PSobs}) and the fit technique (presented in Section \ref{pamfit}) recover the true $f\sigma_8$ from the data, we applied them to the ``average'' measurements from our mock surveys. We expect to recover a value of $f\sigma_8=0.431$ based on the cosmology used to generate our simulations. However, one caveat is that we first need to define an `average' or `typical' power spectrum from the 1000 density and 1000 momentum power spectrum measurements of each survey, which is not trivial if the distribution from the mocks is non-Gaussian. 

Instead, we define the `typical' power in each $k$-bin by using a Gaussian kernel function to smooth the distribution of the measured power spectrum and then identifying the most likely value. In a given $k$-bin, the kernel function is given by \citep{Scott1992,Hartlap2009}
\be\label{Gkenal}
p(x)=\frac{1}{b_wN}\sum^{N=1000}_{i=1}K\left(  \frac{x-P_i(k)}{b_w} \right)
\ee
where the bandwidth $b_w$ is calculated using the ``rule of thumb" ($b_w=3.5\sigma/N^{1/3}$ where $\sigma$ is the sample standard deviation, also known as Scott's rule, \citealt{Scott1992}) and a Gaussian kernel is used for the function $K$\footnote{We used the \textsc{python} function scipy.stats.gaussian\underline{ }kde to compute $p(x)$.}.

\begin{figure}  
\includegraphics[width=\columnwidth]{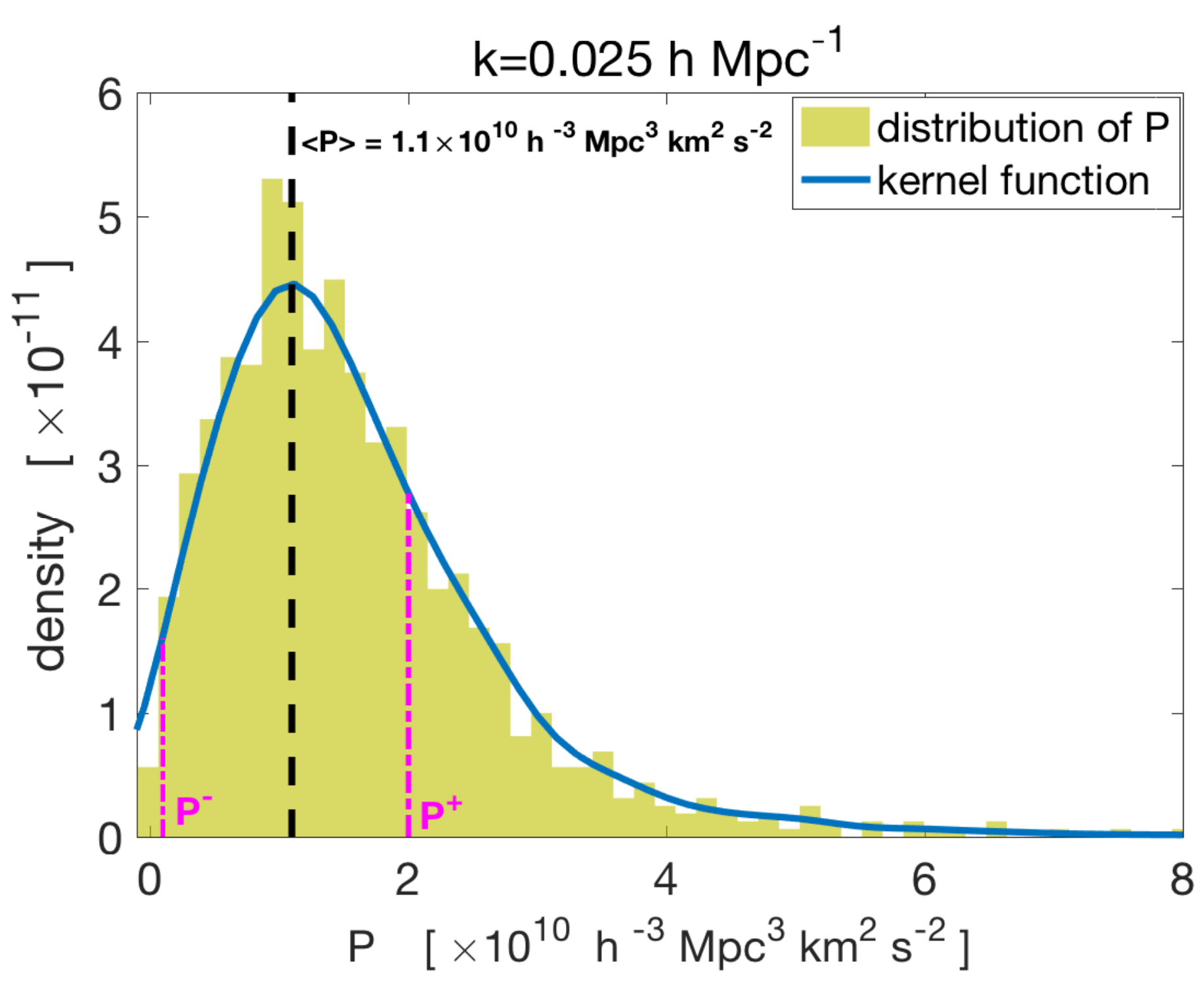}
 \caption{The distribution of the momentum power spectrum measured from $N$=1000 2MTF mocks at $k$=0.025 h$^{-1}$ Mpc. The bars show the normalized histogram of the power spectrum, $P$. The blue curve is the kernel function calculated from $P$ using Eq.~\ref{Gkenal}.The black dashed line indicates the position of the most likely value of the power spectrum. The two pink dashed-dot lines
 at $P^-$ and $P^+$ indicate the boundary of the area for the 68\% confidence level around the most likely value $\langle P \rangle$.}
 \label{KDE}
\end{figure}

As an example, Fig.~\ref{KDE} shows the distribution of the momentum power spectrum measured from $N$=1000 2MTF mocks at $k=0.025h\,\mathrm{Mpc^{-1}}$, in which we also show the smoothed function obtained from our Gaussian kernel function, the maximum likelihood value we use our measurements for testing our fits, and the equal likelihood bounds $P^-$, $P^+$ indicating the 68\% confidence level around the maximum likelihood. We use $P^-$ and $P^+$ to define the error bars of the data point centred at $\langle P \rangle$. It is important to note that for the power spectrum measured from the real 2MTF and 6dFGSv surveys, $P^-$ and $P^+$ are still used to define the error bars used in plotting. However, the `typical' power spectrum measurement is replaced by the power spectrum measured from the data and these error bars are \textit{not} used in the parameter fit. The parameters $f\sigma_8$ and otherwise are fit from the Gaussianized power spectrum $Z$ with covariance matrix calculated from $Z$ as discussed in Section~\ref{pamfit}. To reiterate, the method used here to obtain `typical' power spectrum measurements and errors is used only for testing on the mocks and for plotting error bars on the data points.

\begin{figure}  
\includegraphics[width=\columnwidth]{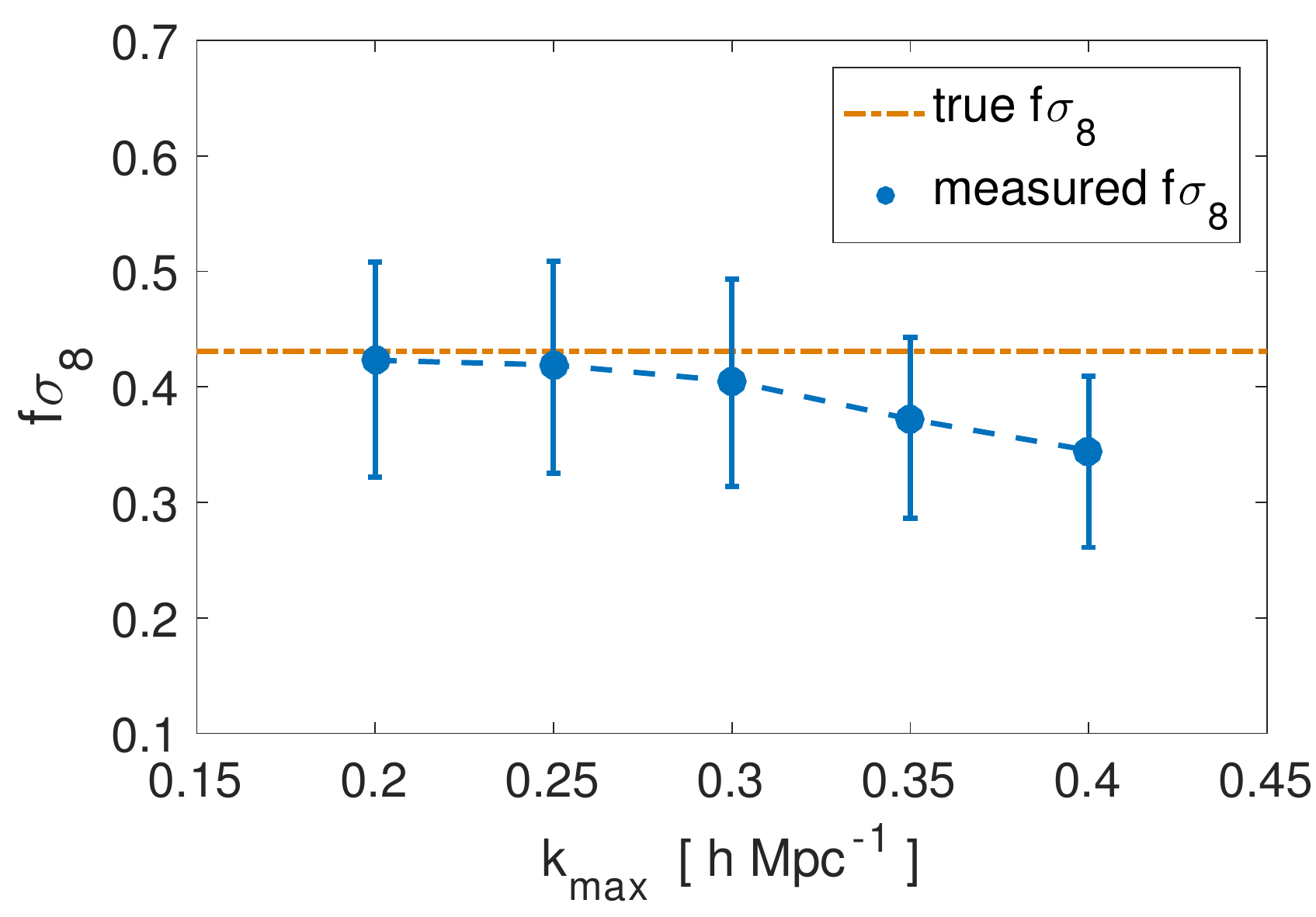}
 \caption{The measured $f\sigma_8$ (blue dots) as a function of the cut-off $k_{max}$ for the combined mocks. The orange dot-dashed line indicates the true $f\sigma_8=0.431$.}
 \label{fkmax}
\end{figure}

Using the combined mocks, we first measure $f\sigma_8$ using the density and momentum power spectrum for $k\in[0.02,~k_{max}]$ in bins of width $0.01h\,\mathrm{Mpc^{-1}}$. Fig.~\ref{fkmax} shows the measured $f\sigma_8$ as a function of the cut-off $k_{max}$ using $k_{max}=$ 0.20, 0.25, 0.30, 0.35 and 0.40 $h$ Mpc$^{-1}$. As we go to higher $k_{max}$ and include more non-linear information in the fit, the measured $f\sigma_8$ is biased away from the true value due to systematic errors in the models of Section~\ref{PSmod}. However, the measurement error of $f\sigma_8$ is also slightly reduced as $k_{max}$ increases. To obtain a good balance between measurement error and systematic bias, we choose $k_{max}=0.3h$ Mpc$^{-1}$. This results in a total $2\times N_{fit}=56$ data points ($28$ for each power spectrum).

\begin{figure}  
\includegraphics[width=\columnwidth]{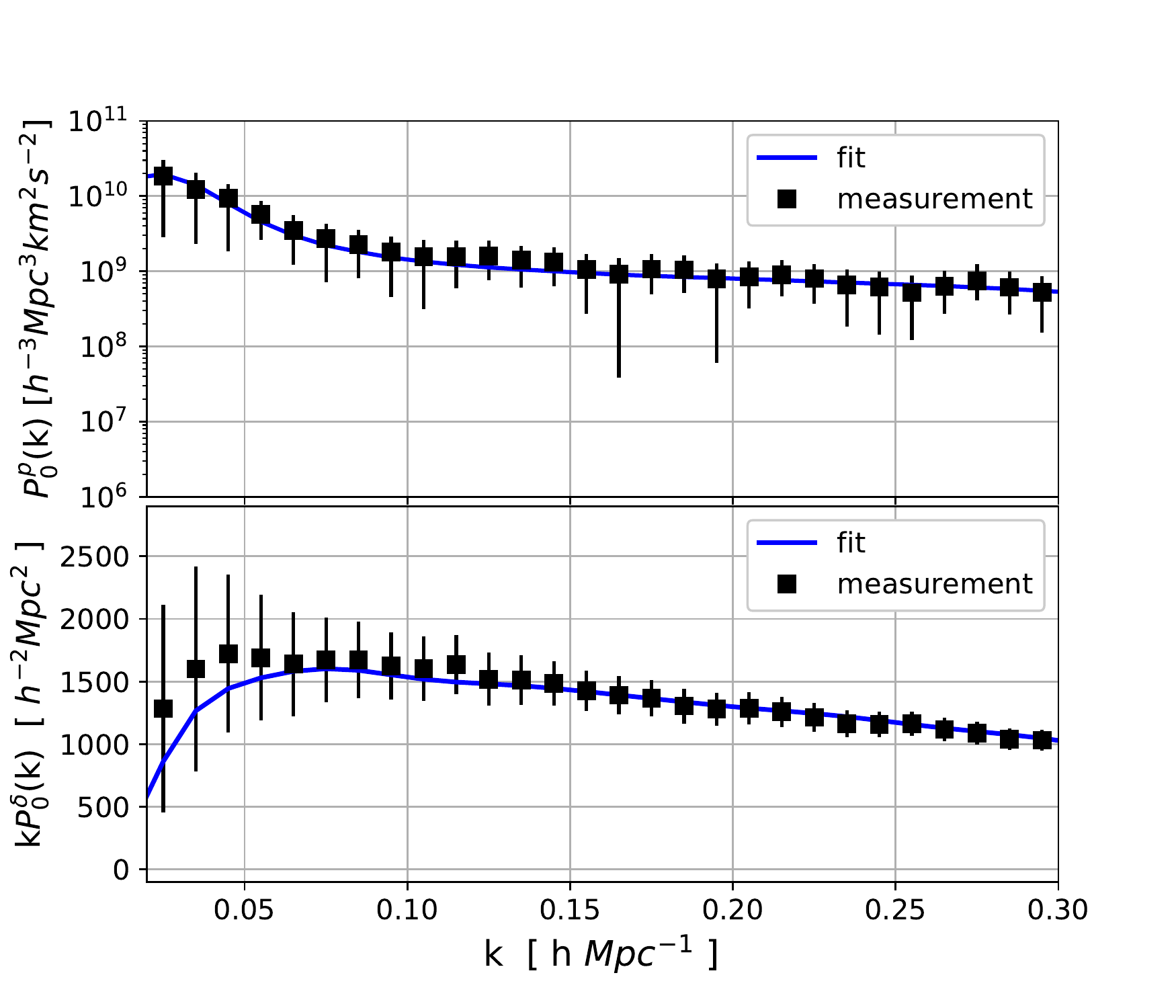}
 \includegraphics[height=73mm,width=73mm]{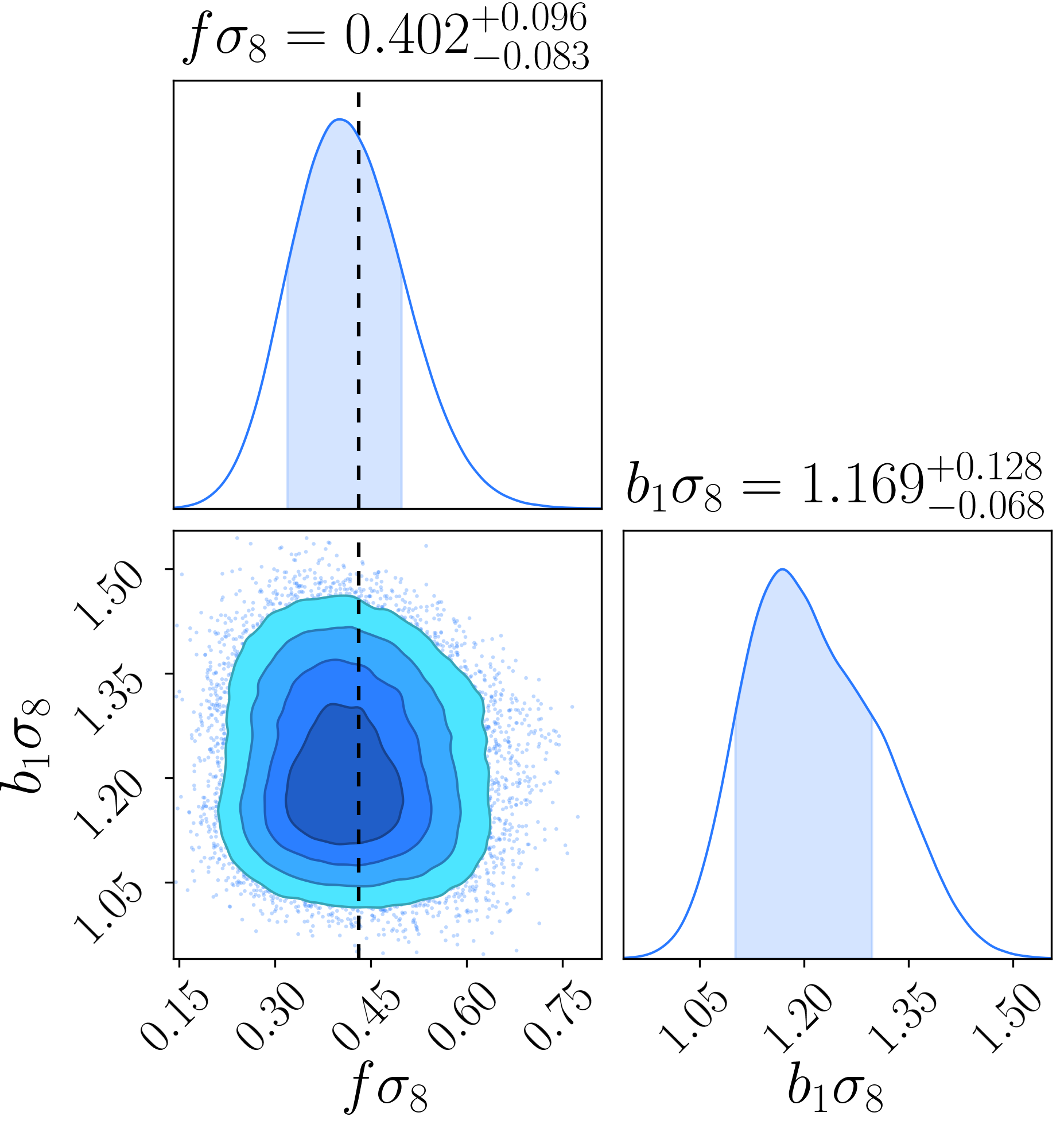}
  \caption{The power spectrum measurements and parameter fit results of the combined mocks. The top panel is the momentum power spectrum. The middle panel is the density power spectrum. The filled squares (with error bars) are the measured power spectrum of the mocks calculated using the kernel function of Eq.~\ref{Gkenal}; the blue curves are the model power spectrum fit to the measurements. The bottom panel shows the 2D contours and the marginalised histograms of the MCMC samples of $f\sigma_8$ and $b_1\sigma_8$, the filled 2D contours indicate the 1, 1.5, 2 and 2.5$\sigma$ regions, whilst the shaded region in the histograms is $1\sigma$. The vertical dashed line indicates the expected value $f\sigma_8=0.431$.}
 \label{cbmocka}
\end{figure}

Fig.~\ref{cbmocka} shows the power spectrum measurements and parameter fit results of the combined 2MTF and 6dFGSv mocks. The recovered cosmological parameters are $f\sigma_{8}=0.402^{+0.096}_{-0.083}$ and $b_1\sigma_8=1.169^{+0.128}_{-0.068}$, as listed in Table \ref{taba1}. Our results are consistent with the expectations of the mocks and unbiased. The results for $f\sigma_8$ and
$b_1\sigma_8$ using the individual 2MTF and 6dFGSv mocks are also shown in Table \ref{taba1}. The error of $f\sigma_8$ measured from the combined mocks has reduced by $\sim18\%$ compared to that of the 2MTF mocks alone and $\sim16\%$ compared to the 6dFGSv mocks alone. This is in agreement with the prediction from \cite{Howlett2017c}.
In Table \ref{taba1},
the $\chi^2/$d.o.f is less than one due to the fact we are fitting the average (or `typical') power spectrum which is close to noise free.

\begin{table}   \centering
\caption{The best fit values and marginalised errors of $f\sigma_8$ and $b_1\sigma_8$ for the 2MTF, 6dFGSv and combined mocks. The number of degrees of freedom is 52 (56 data points and 4 free parameters).}
\begin{tabular}{|c|c|c|c|}
\hline
\hline
Mocks &$f\sigma_8$  &$b_1\sigma_8$ &$\chi^2/$d.o.f \\
\hline
2MTF      & $0.330^{+0.110}_{-0.101}$ &  $0.736^{+0.085}_{-0.068}$ &      24/52 \\ 
\\
6dFGSv    & $0.426^{+0.113}_{-0.094}$ &  $1.098^{+0.087}_{-0.071}$ &    13/52   \\
\\
Combined  & $0.402^{+0.096}_{-0.083}$ &  $1.169^{+0.128}_{-0.068}$ &   11/52 
\\
\hline
\end{tabular}
\label{taba1}
\end{table}

We also constrain $f\sigma_8$, $b_1\sigma_8$ using only the momentum power spectrum from the mocks. The results are shown in Table~\ref{tabs}. The constraints on $b_1\sigma_8$ are weak compared to the values in Table~\ref{taba1} as the momentum power spectrum only contains information on the bias on non-linear scales. However, the errors of $f\sigma_8$ are only $\sim 23\%$ larger than the values in Table~\ref{taba1}; a result of the fact that the momentum power spectrum is proportional to the growth rate of structure on linear scales. Hence we can conclude that the momentum power spectrum is a good tracer of the growth rate of structure on its own, but that the constraints, especially on non-linear scales for 2MTF and 6dFGSv, will be substantially improved by including the density power spectrum.        

\begin{table}   \centering
\caption{The best fit values of $f\sigma_8$ and $b_1\sigma_8$ from fitting the momentum power spectrum only to the mocks. For the 2MTF mocks the parameters are unconstrained (because the noise on non-linear scales allows for a momentum power spectrum consistent with zero) and so we simply quote the 1$\sigma$ upper limit for $f\sigma_8$ and $b_1\sigma_8$. The number of degrees of freedom is 24 (28 data points and 4 free parameters).}
\begin{tabular}{|c|c|c|c|}
\hline
\hline
Mocks &$f\sigma_8$  &$b_1\sigma_8$ &$\chi^2/$d.o.f \\
\hline
2MTF      & $<0.320$ &  $<1.630$ &  5/24 \\
\\
6dFGSv    & $0.309^{+0.126}_{-0.100}$ &  $1.840^{+1.110}_{-0.770}$ &  4/24 \\
\\
Combined  & $0.330^{+0.120}_{-0.100}$ &  $1.490^{+0.920}_{-0.620}$&  5/24  \\
\hline
\end{tabular}
 \label{tabs}
\end{table} 

In this section, we have tested the power spectrum estimator and the parameter fit method on mock surveys. We have verified that our method for measuring, modelling and fitting the density and momentum power spectra works well and should recover the true $f\sigma_8$ from the combined data.

\section{Results and Comparison with GR}\label{fitting}

\subsection{Results}

\begin{figure}  
\includegraphics[width=\columnwidth]{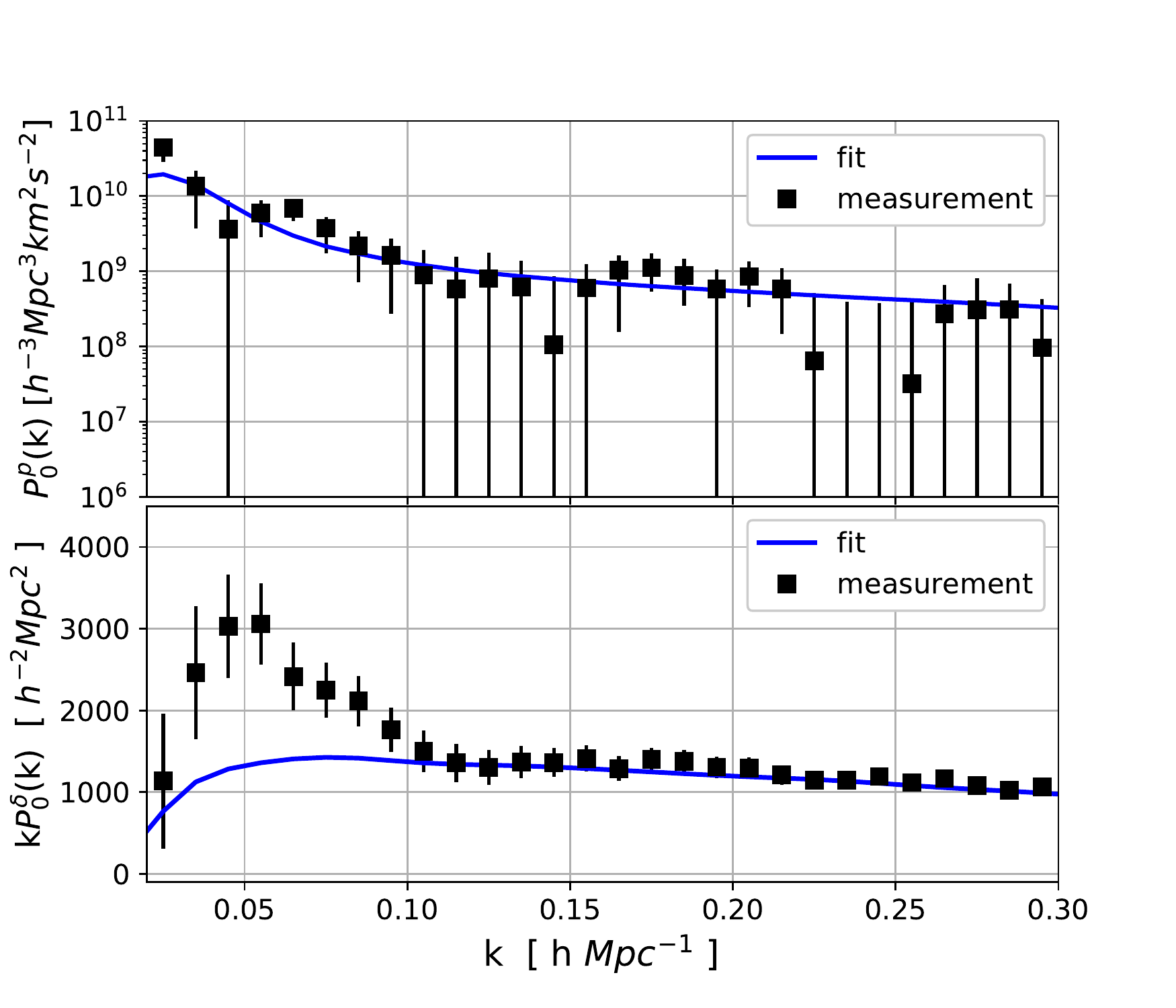}
 \includegraphics[height=73mm,width=73mm]{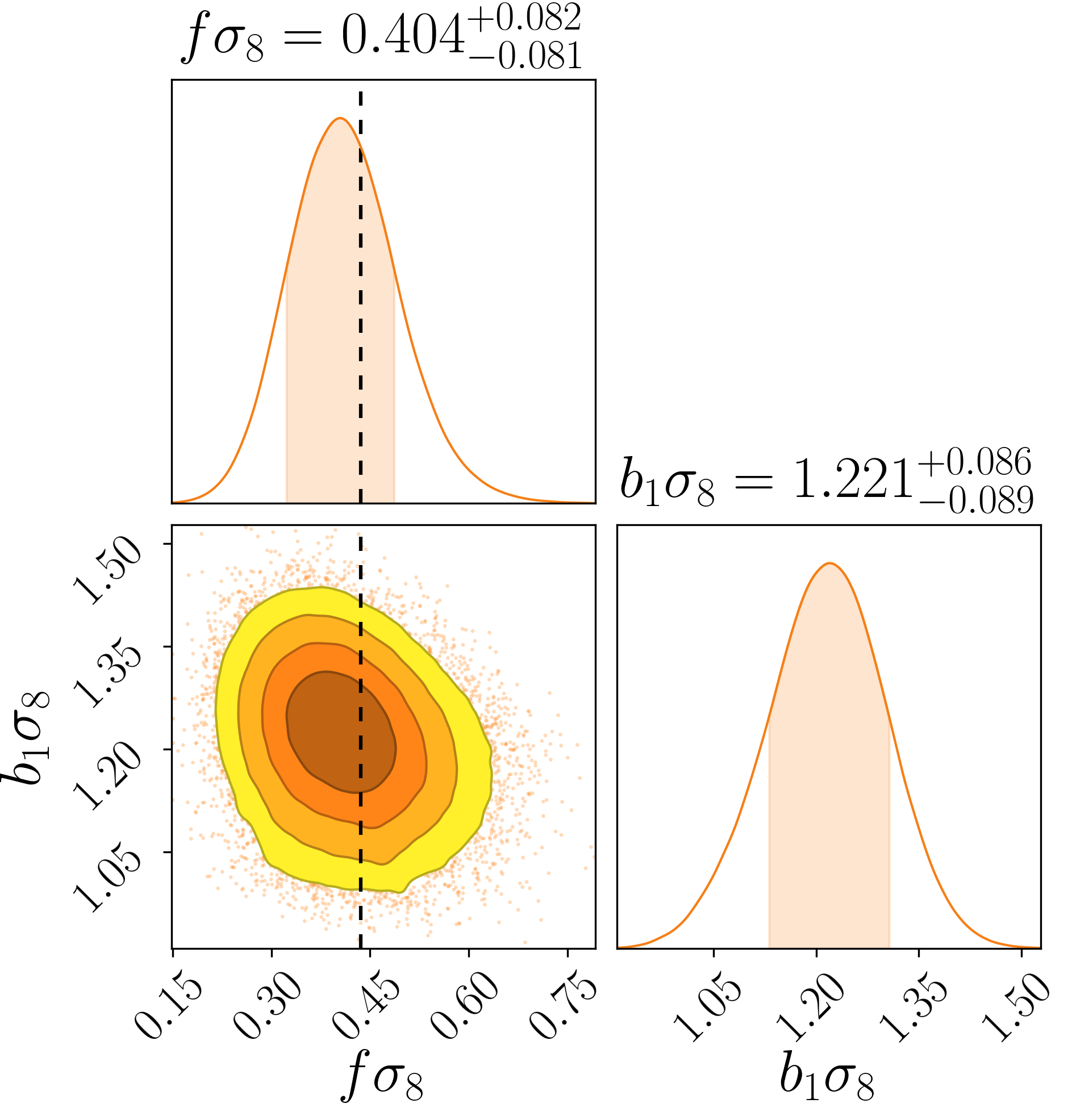} 
\caption{Same as Fig.~\ref{cbmocka} but for the combined 2MTF and 6dFGSv \textit{data}. The vertical dashed line is for the GR prediction $f\sigma_8=0.436$.}
\label{cbpsa}
\end{figure}

We show the power spectrum measurements and parameter fits for the combined 2MTF and 6dFGSv surveys in Fig.~\ref{cbpsa}. Although noisy, we recover a robust measurement of the density and momentum power spectra, and a good fit to the data with a $\chi^{2}$ of 73 for 52 degrees of freedom. The best-fit values for our cosmological parameters of interest are $f\sigma_8=0.404^{+0.082}_{-0.081}$ and 
$b_1\sigma_8=1.221^{+0.086}_{-0.089}$. The recovered accuracy of these parameters is consistent with our fits to the `average' of the combined mocks. Along with measurements from the combined data we also constrain the growth rate using the individual datasets. All our results are shown in Table \ref{taba2}. The measurement error of $f\sigma_8$ measured from the combined surveys is reduced by $\sim 52\%$ and $\sim 23\%$ compared to the individual 2MTF and 6dFGSv surveys, respectively. This is again consistent with what was found using our mocks, and with the predictions of \cite{Howlett2017c}. 
 
\begin{table}   \centering
\caption{The best fit values and marginalised errors of $f\sigma_8$ and $b_1\sigma_8$ for the 2MTF, 6dFGSv and combined data. The number of degrees of freedom is 52 (56 data points and 4 free parameters).}
\begin{tabular}{|c|c|c|c|}
\hline
Surveys &$f\sigma_8$  &$b_1\sigma_8$ &$\chi^2/$d.o.f  \\
\hline
2MTF      & $0.440^{+0.130}_{-0.120}$ &  $0.665^{+0.089}_{-0.074}$ &83/52\\
\\
6dFGSv    & $0.451^{+0.108}_{-0.092}$ &  $1.117^{+0.097}_{-0.086}$ &60/52\\
\\
Combined  & $0.404^{+0.082}_{-0.081}$ &  $1.221^{+0.086}_{-0.089}$&73/52 
\\
\hline
\end{tabular}
\label{taba2}
\end{table}

\subsection{Comparison with theory}

Assuming General Relativity (GR), the growth rate, $f$ is approximately given by \citep{Lahav1991,Strauss1995,Linder2007}
\be 
f(z)\approx\Omega_m(z)^{\gamma},
\ee 
where $\gamma=0.55$ and $\Omega_m(z)$ is the matter density parameter at redshift $z$. Alternative models of gravity can generate deviations away from $\gamma=0.55$, or away from the above parameterisation entirely, introducing additional (possibly scale-dependent) dependencies into the growth rate of structure \citep{Eke1996,Wang1998,Linder2007}. Hence one can use the above model to test the consistency between GR and measurements of the growth rate of structure by fitting for the value of $\gamma$. 

We perform such a test following \cite{Howlett2015}, using the publicly available Planck likelihood chains\footnote{The Planck likelihood chains are downloaded from \url{ https://irsa.ipac.caltech.edu/data/Planck/release_2/ancillary-data/HFI_Products.html}. We used the chains in the file \url{base_plikHM_TTTEEE_lowTEB_lensing_1.txt}.}
to place a prior on $\Omega_{m}(z)$ and to compute the theoretical $f\sigma_8$ as a function of $z$. 
We take into account the fact that different values of $\gamma$ will change both the growth rate of structure and the normalisation of the power spectrum $\sigma_{8}$ using
\be  \label{fsig8th}
f(a)\sigma_8(a)=\Omega_{m}(a)^{\gamma}\sigma_{8,0}\frac{D_{gr}(a*)}{D_{gr,0}}\frac{D_{\gamma}(a)}{D_{\gamma}(a*)},
\ee
where
\be 
a=\frac{1}{1+z}~,~~\Omega_{m}(a)=\frac{\Omega_{m,0}}{a^3E(a)^2},
\ee 
and the growth factors are given by
\be 
\begin{split}
&D_{gr}(a)=\frac{H(a)}{H_0}\int^a_0\frac{da'}{a'^3H(a')^3}, \\
&\frac{D_{\gamma}(a)}{D_{\gamma}(a*)}=\mathrm{exp}\left( \int^a_{a*}\Omega_{m}(a')^{\gamma}dlna' \right).
\end{split}
\ee
The Hubble parameter is given by
\be 
H(a)=H_0\sqrt{\frac{\Omega_{m,0}}{a^3}+\frac{1-\Omega_{m,0}-\Omega_{\Lambda,0}}{a^2}+\Omega_{\Lambda,0}}
\ee 
where $H_0$, $\Omega_{m,0}$ and $\Omega_{\Lambda,0}$ are the Hubble constant, matter density parameter and dark energy density parameter of the present-day Universe. The values for $\sigma_{8,0}$, $\Omega_{m,0}$ and $\Omega_{\Lambda,0}$ are directly read in from the Planck chain. We use $\gamma=$0.55 to compute $f(a)\sigma_8(a)$ from Eq.~\ref{fsig8th} and the corresponding prediction is shown as the green solid curve in Fig.~\ref{comp} alongside our measurements.

In Fig.~\ref{comp}, we also compare the measured $f\sigma_8$, including our $f\sigma_8$ measurements and those from a small selection of recent surveys \citep{Blake2011,Beutler2012,Carrick2015,Howlett2017,Alam2017,Huterer2017,Adams2017,Shi2018,Dupuy2019}, 
to the theoretical prediction. Although these measurements are all largely in agreement with the GR prediction, there is some hint of tension when they are all taken together. To highlight this, we take the above measurements and the Planck chain and perform a fit for $\gamma$. The resulting best fit is $\gamma=0.60\pm0.03$, which is then converted back to a range of $f\sigma_{8}$ values as a function of redshift and shown as the light blue curve in Fig.~\ref{comp}. This highlights the preference in the data for a slightly larger value of $\gamma$, and hence weaker gravitational model. We caution that the exact values for the best-fit and error on $\gamma$ should not be taken too seriously; there is significant overlap (and therefore covariance) between many of the measurements we have used, which we have not accounted for. The measurements cited are also not a complete consensus of all growth rate measurements. Nonetheless, Fig.~\ref{comp} highlights the slight tension between current measurements and GR which may be exacerbated or resolved with the next generation of low redshift galaxy and peculiar velocity surveys such as the Taipan Galaxy Survey \citep{2017PASA...34...47D}, WALLABY \citep{2012PASA...29..359K}, DESI \citep{DESI2016}, SkyMapper \citep{2018arXiv180107834W} and LSST \citep{2008arXiv0805.2366I,Howlett2017b}.

\subsection{Detailed comparison with other measurements}

\begin{figure*}  
 \includegraphics[ width=178mm]{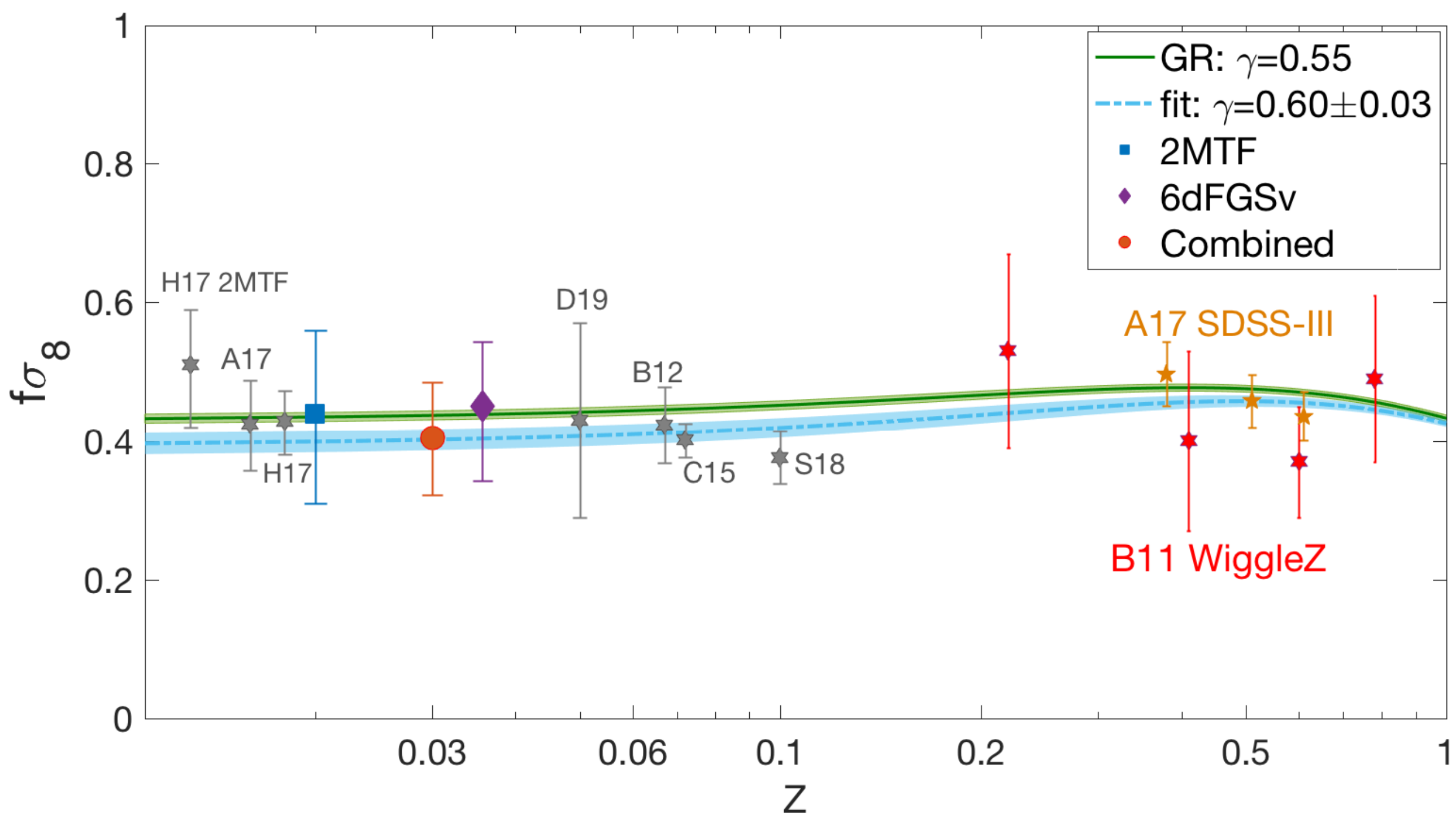}
\caption{Measurements of the growth rate $f\sigma_8$, from the individual and combined 2MTF and 6dFGSv surveys as a function of redshift $z$, compared to measurements from other surveys (stars). 
The blue filled square is the  $f\sigma_8$ measured from the 2MTF survey; the purple filled diamond is from the 6dFGSv survey; and the red filled circle is for the combined surveys. The green solid curve shows the Planck $\Lambda$CDM prediction for $f\sigma_8$ using $\gamma=0.55$; the light-green shaded area indicates the errors. Other measurements are from 
B11 WiggleZ: \citet{Blake2011} using WiggleZ (four red stars);
B12: \citet{Beutler2012} using 6dFGRS;
C15: \citet{Carrick2015} using 2M++, SFI++ and A1; 
H17 2MTF: \citet{Howlett2017} using 2MTF;
A17 SDSS-III: \citet{Alam2017} using SDSS-III (three orange stars);
H17: \citet{Huterer2017}  using JLA+CSP and SN+6dFGSv;
A17: \citet{Adams2017} using 6dFGS; 
S18: \citet{Shi2018} using SDSS-DR7;
D19: \citet{Dupuy2019} using Cosmicflows-3. The light-blue dashed curve is the best fit to the data points, with the light-blue shaded region indicating the errors. The fit corresponds to $\gamma=0.60\pm0.03$. The majority of the individual measurements are in good agreement with the GR prediction, but a combined fit highlights a slight preference for higher $\gamma$ (weaker gravity).}
 \label{comp}
\end{figure*}

Several of the additional measurements shown in Fig.~\ref{comp} use similar datasets to this work and so provide a useful benchmark for comparison.

Firstly, as is the case for our measurement, \cite{Adams2017} use both the density and velocity fields measured using the 6dFGS to constrain $f\sigma_8$. They use comparatively more redshifts in their analysis (20796), adding a large number of galaxies from 6dFGS  that do not have peculiar velocity measurements. They also include a measurement of the cross-correlation between the density and velocity power spectrum and fit to more non-linear scales. All of this leads to smaller error bars compared to our 6dFGSv results which, given they found $\sim20\%$ improvement from including the cross-correlation, is likely in no small part also due to their extra redshifts. However, overall, the best-fit and error on $f\sigma_8$ are still similar and whilst their technique is quite different from ours, the two are highly complementary. Our method offers greater flexibility in the modelling and will likely scale better for large datasets (given that we do not require an integration of the model for every pair of galaxies) whilst their method offers a direct route to extracting the velocity power spectrum. In future, we aim to extend our method to also include the density-momentum cross-correlation and explore any benefit of applying both techniques to the same data and combining the resulting constraints.

Both \cite{Huterer2017} and \cite{Howlett2017} used similar techniques to \cite{Adams2017} but for the velocity field alone. Like \cite{Adams2017}, \cite{Huterer2017} used 6dFGSv to measure $f\sigma_8$, but include an additional 164 Type Ia supernovae (SNIa) into their data set. The total number of SNIa they have used is small, however the precise distance estimates significantly improve the measurement error of $f\sigma_8$, even comparing to \cite{Adams2017}. The methods presented in our paper can also be applied to supernovae from their data set, or from other supernova surveys, such as LSST \citep{Ivezi2008,Howlett2017b}.

\cite{Howlett2017} used the 2MTF dataset. Their error is actually smaller than our measurement even without the inclusion of the density field. This is likely because the small number of galaxies in 2MTF and severe window function mean the information in the density and momentum fields is difficult to extract. For such small datasets the method used in \cite{Howlett2017} is not restricting and we conclude that our technique is probably sub-optimal and loses some information in compressing the data. However, as with the comparison for 6dFGSv, the new method we present here may be preferable if we wished to include more redshifts from the full 2MASS Redshift Survey \citep{Huchra2012}.

\cite{Dupuy2019} recently used the velocity correlation function to constrain $f\sigma_8$ using the CosmicFlows-III dataset. Their data contains significantly more data than used here, but their results are worse because they neglect any information in the density field. This indicates that our method and that of \cite{Adams2017} method are currently more promising, but more work could be done to try and measure and fit the density and velocity correlation functions simultaneously.

Finally, \cite{Carrick2015} used a reconstruction method to fit $f\sigma_8$ obtaining a measurement error that is much smaller than the measurements listed above. Their sample is very large, with 69160 high quality 2M++ galaxies used to reconstruct the density field. 
 Their peculiar velocity catalogue was composed of 2067 field galaxies and 595 galaxy groups. 
They also have 245 SNIa samples in the peculiar velocity catalogue to estimate the growth rate. It is likely that by cleverly reconstructing the density field they are able to include more non-linear information on smaller distance scales than we are able to with the modelling used in this paper. However, as presented in Section 5.1 of \cite{Carrick2015}, there are many  
potential systematic effects that could bias the measurement of $f\sigma_8$ and that are hard to test using simulations. The method we have used in this work allows for more flexible modelling and, as with measurements of redshift-space distortions, can be verified easily up to a well-defined $k_{max}$ scale using mock galaxy catalogues.

\section{conclusion}\label{conc}

In this paper we have presented measurements of the density and momentum power spectrum from the individual and combined 2MTF and 6dFGSv surveys, the first application of the techniques developed in Paper I to real data. We have then used these measurements to estimate the cosmological parameters $f\sigma_8$ and $b_1\sigma_8$ from these surveys. 

We have identified and removed biases that can arise from non-Gaussianity in the measured power spectrum when the survey volume is small, correcting for such effects by applying the Box-Cox transformation to the power spectra to obtain Gaussianized power spectra before fitting. In addition, to account for the scatter of the estimated covariance matrix of the power spectrum, we have used a revised $t$-distribution as the posterior distribution to estimate the parameters.     

We have tested the density and momentum power spectrum estimators (introduced in Section \ref{PSobs}) and our parameter extraction technique (introduced in Section \ref{pamfit}) on the individual and combined 2MTF and 6dFGSv mocks. We find these methods perform well in recovering the true value of $f\sigma_8$ of the simulations. In addition, we have verified that combining the momentum power spectrum with the density power spectrum tightens the constraints on $f\sigma_8$ and $b_1\sigma_8$.  
   
For the 2MTF survey alone we find a value $f\sigma_8=0.440^{+0.130}_{-0.120}$ and $b_1\sigma_8=0.665^{+0.089}_{-0.074}$, whilst for the 6dFGSv survey we find $f\sigma_8=0.451^{+0.108}_{-0.092}$ and $b_1\sigma_8=1.117^{+0.097}_{-0.086}$. Combining these data, we recover $f\sigma_8=0.404^{+0.082}_{-0.081}$ and $b_1\sigma_8=1.221^{+0.086}_{-0.089}$ at an effective redshift of $z_{\mathrm{eff}}=0.03$. These measurements are all consistent with the prediction of General Relativity and the $\Lambda$CDM cosmological model $f\sigma_{8}=0.431$. The measurement error of $f\sigma_8$ measured from the combined 2MTF and 6dFGSv surveys is reduced by $\sim 52\%$ and $\sim 23\%$ compared to that of the individual 2MTF and 6dFGSv surveys, respectively. This is agreement with the prediction of \cite{Howlett2017c}.

\section*{Acknowledgements}
Fei Qin has received financial support from the China Scholarship Council (CSC). Parts of this research were conducted by the Australian Research Council Centre of Excellence for All-sky Astrophysics (CAASTRO), through project number CE110001020 and the Australian Research Council Centre of Excellence for All Sky Astrophysics in 3 Dimensions (ASTRO 3D), through project number CE170100013. This research has made use of the \textsc{ChainConsumer} package \citep{ChainConsumer}.



\bibliographystyle{mnras}
\bibliography{BKFmn}



\appendix

\section{2MTF and 6\lowercase{d}FGS\lowercase{v} mock catalogues}\label{sec:Ap1}

In this work we use a set of 1000 mock 2MTF and 6dFGSv surveys to evaluate our covariance matrix. Producing such a large ensemble of mock surveys in a feasible time requires approximate N-Body methods as opposed to fully non-linear simulations. We use the \textsc{l-picola} code \citep{Howlett2015bs,Howlett2015cs} to generate 125 $z=0$ dark matter simulations with the fiducial cosmology given in Section~\ref{sec:introduction} but with different initial conditions. Halos (but not subhalos) were identified in these simulations using the 3D Friends of Friends algorithm included in the \textsc{VELOCIraptor} code \citep{Elahi2009}. Each of the simulations contains $2560^3$ particles in a box of length $L=1800h^{-1}\mathrm{Mpc}$ and a minimum of 20 particles per halo was required, leading to a minimum halo mass of $\sim5\times10^{11}M_{\odot}h^{-1}$. 

As shown in \cite{Howlett2015cs}, the approximate nature of the simulations compared to a fully non-linear N-body simulation leads to a suppression of non-linear clustering and a reduction in the number of low-mass halos. However, by comparing \textsc{l-picola} and fully-nonlinear (\textsc{gadget-2}; \citealt{Springel2005ga}) simulations generated with the same initial conditions, these same works demonstrated that the dark matter clustering is recovered to within $1\%$ up to $k=0.3h\,\mathrm{Mpc^{-1}}$, whilst the cross-correlation coefficient (which can be treated as an indicator of how well the covariance matrix is recovered) is accurate to even smaller scales. \cite{Howlett2015} also found that the lack of low mass halos can be easily accounted for when galaxies are assigned to the halos; such an assignment typically contains enough flexibility that one can still recover the same galaxy clustering but the parameters used for the modelling will be slightly different between \textsc{l-picola} and \textsc{gadget-2} simulations.

In previous work \citep{Howlett2017,Qin2018a,Qin2019} we used Subhalo Abundance Matching \citep{Conroy2006} to assign galaxies to the dark matter simulations before applying the selection functions of the 2MTF or 6dFGSv data. A key difference in this work is that we do not expect the \textsc{l-picola} simulations to accurately reproduce subhalos and so we require an additional step that assigns substructures to the parent halos and allows for the possibility that these surveys contain both central and satellite populations before the galaxy luminosities or Fundamental Plane parameters are assigned.

Motivated by the fact that the subhalo mass function is largely characterised by a power law \citep{Springel2008, Giocoli2008, Elahi2018}\footnote{Here we neglect the effects of stripping of the subhalos as they fall into their parent halo, and the presence of long-lived major merger remnants which can cause an upturn in the number of subhalos with large masses compared to their parent halos}, we write the expected number of subhalos $N_{sub}$ as a function of the mass ratio, $f_{M}$ between the subhalo and its parent,
\begin{equation}
    N_{sub}(f_{M}) = Af_{M}^{-\alpha} .
\end{equation}
The total number of subhalos in each halo is then given by integrating this function between the minimum subhalo mass fraction $f_{min}$ (which is calculated by fixing the minimum subhalo mass to 20 particles) and 1. Integrating this function is equivalent to calculating the CDF, and so once the total number of subhalos is known, we can inverse sample this to generate the individual masses. In practice, we allow for scatter in the total number of subhalos at fixed parent halo mass by drawing the total number of subhalos from a Poisson distribution with mean $\int_{f_{min}}^{1} N_{sub}(f_{M})$, before sampling this many objects from the true power law distribution.

Once we have generated a number of subhalo masses for each halo, these are distributed in the halo assuming an NFW profile \citep{Navarro1997} with mass-concentration relation given by \cite{Prada2006}\footnote{We generate the positions of the subhalos by drawing from the inverse CDF of the NFW profile \citep{Robotham2018}, and use the virial theorem to determine the satellite velocity as a function of position \citep{Lagos2018}}. Once all subhalos have been generated we can perform the same subhalo abundance matching technique to generate mock 2MTF and 6dFGSv surveys as was used in \cite{Qin2018a}. In order to include scatter in the observed halo-mass/luminosity relationship, we assign log-luminosities to subhalos based on a `matching' log-luminosity which is drawn from a Gaussian centred on the true log-luminosity with some standard deviation, $\sigma_{logL}$. Note that this is \textit{only} used for the matching; the luminosity given to each subhalo is still the true luminosity drawn from the luminosity function for 2MTF or calculated from the 6dFGSv Fundamental Plane parameters.

Overall, the above procedure involves three free parameters for each survey, the normalisation and index for the subhalo mass function $A$ and $\alpha$, and the scatter in the mass/luminosity relationship $\sigma_{logL}$, all of which will change the overall clustering of the mock catalogues. For our estimation of the covariance matrix, the most important aspect of this procedure is that the linear bias of the mocks (or rather the amplitude of the clustering) matches that of the data. In order to achieve this we fit the above parameters such that the observed density power spectrum of the mocks matches the data after various selection functions are applied. It is important to note that we do not fit the momentum power spectrum in this process. The amplitude of the momentum power spectrum is proportional to the growth rate, and we do not \textit{a priori} assume that the growth rate of the real Universe matches that of the mocks. 

The fit was done by iterating over the three free parameters by brute force. We read 8 halo catalogues, assigned subhalos, produced sets of mock galaxies, applied selection functions and then measured the average redshift-space density power spectrum of the mocks before computing the chi-squared value with respect to the data. This process requires an estimate of the covariance matrix to begin with, so the whole fitting process itself was performed iteratively three times to ensure that using an updated covariance matrix did not change the results of the fit.

The fitting procedure was performed separately for the 2MTF and 6dFGSv mocks. The best-fit parameters for the two samples are given in Table~\ref{tab:mockparams}. The 2MTF parameters were also used to produce a set of simulations representative of what could be achieved with a future Type-IA supernovae peculiar velocity sample (i.e., following the specifications of \citealt{Howlett2017b,Kim2019}), which were presented in Paper I. 

\begin{table}   \centering
\caption{Best-fit values used for assigning subhalos and luminosities to \textsc{l-picola} halo catalogues.}
\begin{tabular}{|c|c|c|c}
\hline
\hline
Mocks &$A$  &$\alpha$ &$\sigma_{logL}$ \\
\hline
2MTF      & $1.265$ &  $0.743$ & $0.260$\\
6dFGSv      & $1.558$ &  $1.580$ & $0.136$\\
\hline
\end{tabular}
 \label{tab:mockparams}
\end{table} 

For each simulation we place 8 individual observers in the box, maximally far apart, leading to $1000$ mock Universes. The same observers were used to create the 2MTF and 6dFGSv mocks, such that mocks for the combined sample were created by simply adding the two mock surveys for each observer together. The volume of the simulation box and the 2MTF and 6dFGSv surveys is such that the shortest distance between two observers drawn from the same simulation is $\sim600h^{-1}\mathrm{Mpc}$, so we expect the fact that we have drawn multiple observers from the same box to have negligible impact on our estimate of the covariance matrix.

\section{the intrinsic variance of the peculiar velocity field}\label{intri}
\begin{figure}  
\includegraphics[width=\columnwidth]{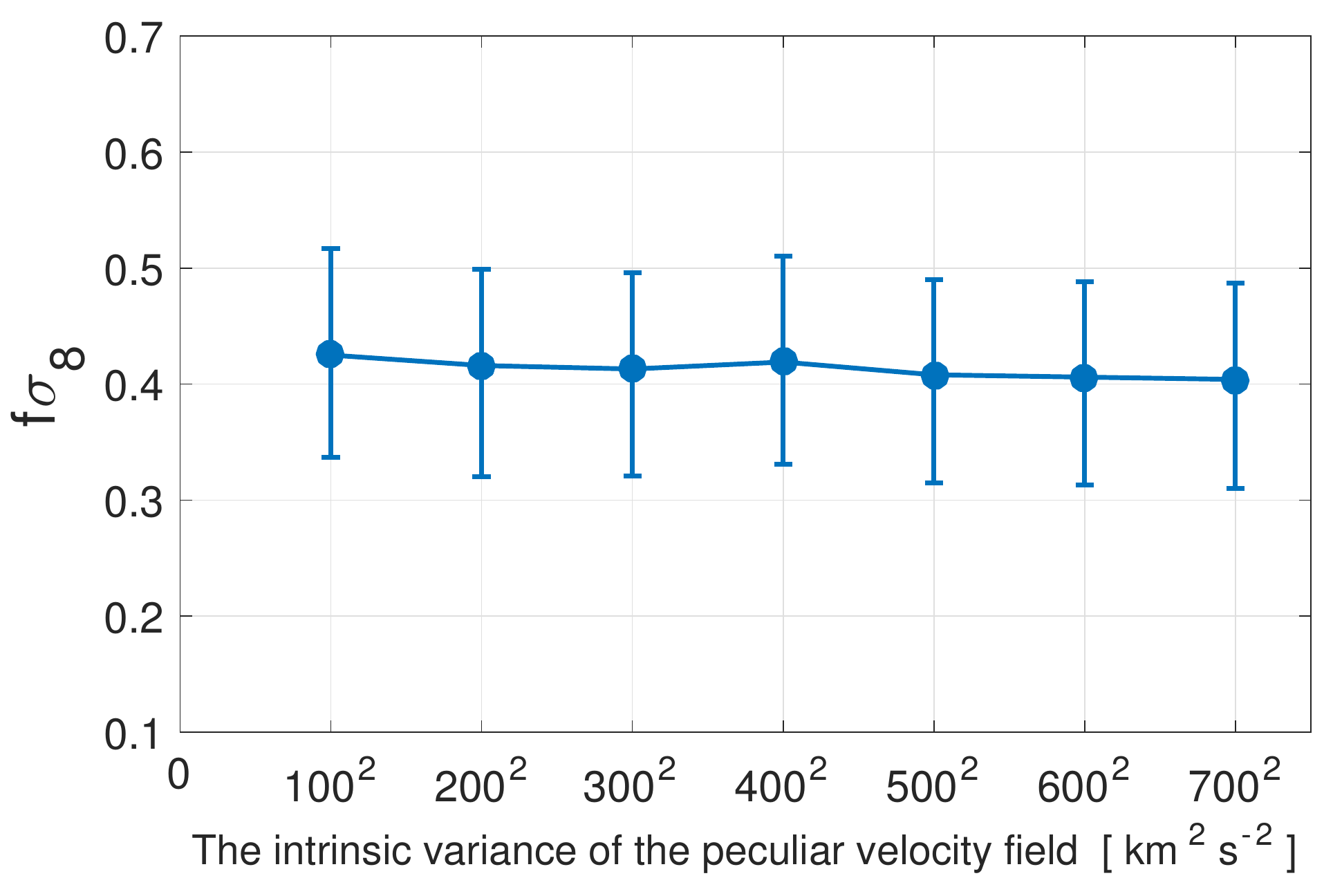}
 \caption{The measured $f\sigma_8$ as a function of the intrinsic variance of the peculiar velocity field. Using the combined 2MTF and 6dFGSv mocks.}
 \label{iss}
\end{figure}

In previous peculiar velocity field measurements, the intrinsic variance of the peculiar velocity field is usually assumed to be $250^2\sim300^2$ km$^2$ s$^{-2}$ \citep{Jaffe1995,Sarkar2007,Feldman2010,Kashlinsky2010,Dai2011,Turnbull2012,Hong2014,Scrimgeour2016,Qin2018a}. The value of this intrinsic variance affects our
results only very weakly.
As an example, we used the combined 2MTF and 6dFGSv mocks to test a variety of values for the intrinsic variance between $100^2\,\mathrm{km^{2}\,s^{-2}}$ and $700^2\,\mathrm{km^{2}\,s^{-2}}$ and found negligible change in the best-fit values and errors on the recovered growth rate, as shown in Fig.\ref{iss}. 
 
\bsp	
\label{lastpage}
\end{document}